\documentclass{article}
\usepackage[dvips]{graphicx}
\usepackage{graphicx}

\newcommand{\nc}{\newcommand}
\nc{\noi}{\noindent}  \nc{\Nrm}{{\rm N}}
\nc{\erm}{{\rm e}}  \nc{\Rcal}{{\cal R}}
\nc{\AR}{A_{\rm 1R}} \nc{\ARAR}{A_{\rm 2R}}
\nc{\AT}{A_{\rm 1T}} \nc{\ATAT}{A_{\rm 2T}}
\nc{\al}{\alpha_1} \nc{\alal}{\alpha_2}
\nc{\be}{\beta_1} \nc{\bebe}{\beta_2}
\nc{\psiuno}{\psi_{\rm I}} \nc{\psidue}{\psi_{\rm II}}
\nc{\psitre}{\psi_{\rm III}} \nc{\psiquattro}{\psi_{\rm IV}}
\nc{\psicinque}{\psi_{\rm V}}    \nc{\srm}{{\rm s}}
\nc{\bb}{\begin{equation}} \nc{\ee}{\end{equation}}
\nc{\um}{{1\over 2}} \nc{\C}{I\!\!\!C} \nc{\R}{I\!\!R}
\nc{\pa}{\partial} \nc{\ug}{\; = \;}
\nc{\cent}{\centerline}

 \newcommand{\ebf}{\mbox{\boldmath $e$}}

 \newcommand{\xbf}{\mbox{\boldmath $x$}}
 
 \newcommand{\zbf}{\mbox{\boldmath $z$}}

 \newcommand{\imp}{\mbox{\boldmath $p$}}
 \newcommand{\Ebf}{\mbox{\boldmath $E$}}
 \newcommand{\kbf}{\mbox{\boldmath $k$}}
 
 \newcommand{\Hbf}{\mbox{\boldmath $H$}}
 \newcommand{\rbf}{\mbox{\boldmath $r$}}

 \newcommand{\jbf}{\mbox{\boldmath $j$}}
 \newcommand{\Abf}{\mbox{\boldmath $A$}}
 \newcommand{\chibf}{\mbox{\boldmath $\chi$}}

\begin{document}

\title{On non-selfadjoint operators for observables in quantum mechanics \\
and quantum field theory}

\author{Erasmo~Recami$^1$\footnote{E-mail: recami@mi.infn.it},
Vladislav~S.~Olkhovsky$^2$\footnote{E-mail: olkhovsk@kinr.kiev.ua}  and
Sergei~P.~Maydanyuk$^2$\footnote{E-mail: maidan@kinr.kiev.ua}\\
\small{\em $^1$ Facolt\`{a} di Ingegneria, Universit\`{a} statale di Bergamo,
Bergamo, Italy {\rm and} INFN-Sezione di Milano, Milan, Italy}\\
\small{\em $^2$ Institute for Nuclear Research, National Academy of Science of Ukraine, %\newline
47 Prosp. Nauki, Kiev-03680, Ukraine}}

%%%\begin{document}

\maketitle

\begin{abstract}
Aim of this paper is to show the possible significance, and usefulness,
of various non-selfadjoint operators for suitable Observables in non-relativistic
and relativistic quantum mechanics, and in quantum electrodynamics.
More specifically, this work starts dealing with: (i) the maximal
hermitian (but not selfadjoint) {\em Time} operator in non-relativistic quantum
mechanics and in quantum electrodynamics; and with: (ii) the problem of the
four-position and four-momentum operators, each one with its hermitian and
anti-hermitian parts, for relativistic spin-zero particles.  \ Afterwards,
other physically important applications of non-selfadjoint (and even
non-hermitian) operators are discussed: In particular, (iii) we reanalyze in
detail the interesting possibility of associating quasi-hermitian
Hamiltonians with (decaying) {\em unstable states} in nuclear physics.  Finally,
we briefly mention the cases of quantum dissipation, as well as of the nuclear
optical potential.

\end{abstract}

\

\

{\bf PACS numbers:}
03.65.Ta; 03.65.-w; 03.65.Pm; 03.70.+k; 03.65.Xp; 11.10.St; 11.10.-z; 11.90.+t; 02.00.00; 03.00.00; 
24.10.Ht; 03.65.Yz; 21.60.-u; 11.10.Ef; 03.65.Fd 

\

\

{\bf Keywords:} time operator, space-time operator,
time-``Hamiltonian", non-selfadjoint operators, non-hermitian
operators, bilinear operators, time operator for discrete energy
spectra, time-energy uncertainty relations, unstable states,
quasi-hermitian hamiltonians, Klein-Gordon equation, quantum
dissipation, nuclear optical model
% *******************************************************************************************************************

% *******************************************************************************************************************
\newpage
+-\tableofcontents
% *******************************************************************************************************************

% *******************************************************************************************************************
\newpage
\section{Introduction
\label{sec.Introduction}}

Time, as well as 3-position, sometimes is a parameter, but sometimes is an onservable that
in quantum theory would be expected to be associated with an operator. However, almost from
the birth of quantum mechanics (cf., e.g., Ref.\cite{Pauli.1926}), %[1]),
it is known that time cannot be represented by a selfadjoint operator, except in the case of
special systems
(such as an electrically charged particle in an infinite uniform electric field)%
\footnote{This is a consequence of the semi-boundedness of the continuous energy spectra
from below (usually from zero).
Only for an electrically charged particle in an infinite uniform electric field, and other
very rare special systems,
the continuous energy spectrum is not bounded and extends over the whole axis from
$-\infty$ to $+\infty$. It is curious
that for systems with continuous energy spectra bounded from above and from below, the time
operator is however selfadjoint and yields a discrete time spectrum.}.
The list of papers devoted to the problem of time in quantum mechanics is extremely large (see, for
instance,
Refs.[2--29], and references therein). The same situation had to be faced  also in quantum
electrodynamics and, more in
general, in relativistic quantum field theory (see, for instance,
Refs.\cite{Olkhovsky_Recami.1968.NuovoCim,PhysRep2004,IJMPA}). %[2, 17]).

As to quantum mechanics, the very first relevant articles are probably Refs.[2--9], and
refs. therein. \ A second set of papers on time in
quantum physics[10--29] appeared in the nineties, stimulated partially by the need of a
consistent definition for the
tunneling time. It is noticeable, and let us stress it right now, that this second set of
papers seems however to have ignored Naimark's theorem\cite{Naimark.1940}, %[28],
which had previously constituted (directly or indirectly) an important basis
for the results in Refs.\cite{Olkhovsky_Recami.1968.NuovoCim}--\cite{Holevo.1978.RMP}; %[2--8]
moreover, all the papers\cite{Srinivas.1981.PJP}--\cite{Atmanspacher.1998.IJTP} % [9-16]
attempted at solving the problem of time as a quantum observable by means of formal
mathematical operations performed {\em outside} the usual Hilbert
space of conventional quantum mechanics. \
Let us recall that Naimark's theorem states\cite{Naimark.1940} %[28]
that the non-orthogonal spectral decomposition of a hermitian operator
{\em can be approximated} by an orthogonal spectral function (which corresponds to a
selfadjoint operator), in a weak convergence, {\em with any desired
accuracy}.

The main goal of the first part of the present paper is to justify the use of  time
as a quantum observable,
basing ourselves on the properties of the hermitian (or, rather, maximal hermitian)
operators for the case of continuous
energy spectra: cf., e.g., the
Refs.\cite{Olkhovsky_Recami.1992PR-1995JP-2004PR-2007IJMP,PhysRep2004,IJMPA}). %[17]).

The question of time as a quantum-theoretical observable is conceptually connected with
the much more general problem
of the four-position operator and of the canonically conjugate four-momentum operator, both
endowed with an hermitian and
an anti-hermitian part, for relativistic spin-zero particles: This problem is analyzed in
the second part of the present paper.

In the third part of this work, it is shown how non-hermitian operators can be
meaningfully and extensively
used, for instance, for describing {\em unstable states} (decaying resonances). \ Brief mentions
are added of the cases of quantum dissipation, and of the nuclear optical potential.

%-----------------------------------------------------------------------------------------------------------------------

%-----------------------------------------------------------------------------------------------------------------------
\section{Time operator in non-relativistic quantum mechanics and in quantum electrodynamics
\label{sec.2}}

\subsection{On {\em Time} as an Observable in non-relativistic quantum mechanics for systems
with continuous energy spectra
\label{sec.2.1}}

The last part of the above-mentioned list[11--29] of papers, in particular Refs.[12-17,22-29]
appeared in the nineties, devoted to the problem of Time in non-relativistic quantum
mechanics, essentially because of the need to define the tunnelling
time.
%\cite{Leon.1997.JPA,Olkhovsky_Recami.1992PR-1995JP-2004PR-2007IJMP,Delgado.1999.PRA,Muga.1999,PhyRep2004,IJMPA}. \
%[14, 17, 22, 23].
As remarked, those papers did not refer to the Naimark theorem
\footnote{The Naimark theorem states in particular the following\cite{Naimark.1940}: %[28]:
The non-orthogonal spectral decomposition of a maximal hermitian operator can be approximated by an orthogonal spectral
function (which corresponds to a selfadjoint operator), in a weak convergence, with any desired
accuracy.} \cite{Naimark.1940}%[28],
which had mathematically supported, on the contrary, the results in [2--9], and afterwards
in [18-21]. %%% and the review \cite{Delgado.1999.PRA}.???%[22]

Indeed, already in the seventies (in Refs.[2--6], while more detailed
presentations and reviews can be found
in\cite{Olkhovsky.1984.SJPN,Olkhovsky.1990.Nukleonika} %[6, 7])
 and independently in~\cite{Holevo.1978.RMP}), %[8],
it was proven that, for systems with continuous energy spectra, Time {\bf is} a quantum-mechanical observable, canonically
conjugate to energy. Namely, it had been shown the time operator
\begin{equation}
  \hat{t} =
  \left\{
  \begin{array}{cll}
    t, & \mbox{in the time ($t$-)representation}, & \mbox{(a)} \\
    -i\hbar\, \displaystyle\frac{\partial}{\partial E}, & \mbox{in the energy ($E$-)representation} & \mbox{(b)}
  \end{array}
  \right.
\label{eq.2.1.1}
\end{equation}
to be not selfadjoint, but hermitian, and to act on square-integrable space-time wave packets
in the representation (\ref{eq.2.1.1}a), and on their Fourier-transforms in (\ref{eq.2.1.1}b),
once point $E=0$ is eliminated (i.e., once one deals only with
moving packets, excluding any {\em non-moving} rear tails and the cases with zero fluxes)%
\footnote{Such a condition is enough for operator (\ref{eq.2.1.1}a,b) to be a {\em hermitian},
or more precisely a {\em maximal hermitian}[2--8] {\em operator}  \ (see
also~\cite{Olkhovsky_Recami.1992PR-1995JP-2004PR-2007IJMP,PhysRep2004,IJMPA,
Olkhovsky.1997.conf}); but it can be dispensed with by recourse to bilinear forms
(see, e.g., Refs.\cite{Recami.1976,Recami.1983.HJ}
and refs. therein), as we shall see below.}
% [17, 18, 29]).}.
%
In Refs.\cite{Olkhovsky.1984.SJPN,Olkhovsky.1990.Nukleonika} and [18-21] %[6, 7, 22]
the operator $\hat{t}$  (in the $t$-representation) had the
property that any averages over time, in the one-dimensional (1D)
scalar case, were to be obtained by use of the following {\em
measure} (or weight):
\begin{equation}
  W\,(t,x)\: dt = \displaystyle\frac{j\,(x,t)\, dt}{\int\limits_{-\infty}^{+\infty} j\,(x,t)\, dt} \; ,
\label{eq.2.1.2}
\end{equation}
where the the flux density $j\,(x,t)$ corresponds to the (temporal) probability for a particle
to pass through point $x$ during the unit time centered at $t$,
when traveling in the positive $x$-direction.  Such a measure is
not postulated, but is a direct consequence of the well-known
probabilistic {\em spatial} interpretation of $\rho\,(x,t)$
 and of the continuity relation $\partial\rho\,(x,t) /
\partial\, t + {\rm div} j\,(x,t) = 0$.  Quantity $\rho(x,t)$ is,
as usual, the probability of finding the considered moving
particle inside a unit space interval, centered at point $x$, at
time $t$.

Quantities $\rho(x,t)$ and  $j\,(x,t)$ are related to the wave function $\Psi\,(x,t)$ by the
ordinary definitions $\rho\,(x,t)=|\Psi\,(x,t)|^{2}$ and $j\,(x,t) = \Re [\Psi^{*}(x,t)\:
(\hbar/i\mu)\, \Psi\,(x,t))]$).  When the flux density $j\,(x,t)$ changes its
sign, quantity $W\,(x,t)\,dt$ is no longer positive-definite and, as in
Refs.\cite{Olkhovsky.1984.SJPN,Olkhovsky_Recami.1992PR-1995JP-2004PR-2007IJMP,PhysRep2004,IJMPA,Olkhovsky.1997.conf}, % [6, 17, 18],
it acquires the physical meaning of a probability density {\em only} during those partial
time-intervals in which the flux density $j\,(x,t)$ does keep its sign. Therefore,
let us introduce the {\em two} measures\cite{Olkhovsky_Recami.1992PR-1995JP-2004PR-2007IJMP,PhysRep2004,IJMPA} %
by separating the positive and the negative flux-direction values (that is, the flux signs)

\begin{equation}
  W_{\pm}\,(t,x)\: dt = \displaystyle\frac{j_{\pm}\,(x,t)\, dt}{\int\limits_{-\infty}^{+\infty} j_{\pm}\,(x,t)\, dt}
\label{eq.2.1.3}
\end{equation}
with $j_{\pm}\, (x,t) = j\,(x,t)\, \theta (\pm j)$.

Then, the mean value $\langle t_{\pm} (x) \rangle$ of the time $t$ at which the particle passes
through position $x$, {\em when traveling in the positive or
negative direction}, is, respectively,
\begin{equation}
  \langle t_{\pm} (x) \rangle =
  \displaystyle\frac{\displaystyle\int\limits_{-\infty}^{+\infty} t\,j_{\pm}\,(x,t)\: dt}
    {\displaystyle\int\limits_{-\infty}^{+\infty} j_{\pm}\,(x,t)\: dt} =
  \displaystyle\frac{\displaystyle\int\limits_{0}^{+\infty}
    \displaystyle\frac{1}{2}\,
    \Bigl[ G^{*}(x,E)\, \hat{t}\, v\, G\,(x,E) +  v\, G^{*}(x,E)\, \hat{t}\, G\,(x,E) \Bigr] \: dE}
    {\displaystyle\int\limits_{0}^{+\infty} v\, \bigl|G\,(x,E)\bigr|^{2} \: dE} \; ,
\label{eq.2.1.4}
\end{equation}
where $G\,(x,E)$ is the Fourier-transform of the moving 1D
wave-packet
\[
\begin{array}{ccl}
  \Psi\, (x,t) & = &
  \displaystyle\int\limits_{0}^{+\infty} G\,(x,E)\, \exp(-iEt/\hbar)\: dE = \\
  & = & \displaystyle\int\limits_{0}^{+\infty} g(E)\, \varphi(x,E)\, \exp(-iEt/\hbar)\: dE
\end{array}
\]
when going on from the time to the energy representation. For free motion, one has
$G(x,E)=g(E)\,\exp(ikx)$, and $\varphi(x,E)=
\exp(ikx)$,  while $E= \mu\, \hbar^{2} k^{2}/\,2= \mu\, v^{2}/\,2$.
In Refs.\cite{Olkhovsky_Recami.1992PR-1995JP-2004PR-2007IJMP,PhysRep2004,IJMPA}, there were defined
the mean time {\em durations} for the particle 1D transmission from $x_{i}$ to $x_f > x_{i}$, and
reflection from the region ($x_{i}$, $+\infty$) back to the interval $x_{f} \le x_{i}$. \ Namely % [17]:
\begin{equation}
  \langle \tau_{T} (x_{i}, x_{f}) \rangle = \langle t_{+} (x_{f}) \rangle - \langle t_{+} (x_{i}) \rangle
\label{eq.2.1.6}
\end{equation}
% (3a)
and
\begin{equation}
  \langle \tau_{R} (x_{i}, x_{f}) \rangle = \langle t_{-} (x_{f}) \rangle - \langle t_{+} (x_{i}) \rangle
\label{eq.2.1.7},
\end{equation}
% (3b)
respectively. \ The 3D generalization for the mean durations of quantum collisions
and nuclear reactions appeared in~\cite{Olkhovsky.1984.SJPN,Olkhovsky.1990.Nukleonika}. % [6, 7].
 \ Finally, suitable definitions of the averages $\langle t^{n} \rangle$ on time
of $t^{n}$, with $n=1,2\ldots$, and of $\langle f(t) \rangle$, quantity $f(t)$
being any analytical function of time, can be found in~\cite{IJMPA,IJMPB}, where
single-valued expressions have been explicitely written down.

The two canonically conjugate operators, the time operator (\ref{eq.2.1.1}) and the energy
operator%*)
\begin{equation}
  \hat{E} =
  \left\{
  \begin{array}{cll}
    E, & \mbox{in the energy ($E$-) representation}, & \mbox{(a)} \\
    i\hbar\, \displaystyle\frac{\partial}{\partial t}, & \mbox{in the time ($t$-) representation} & \mbox{(b)}
  \end{array}
  \right.
\label{eq.2.1.13}
\end{equation}
do clearly satisfy the commutation relation\cite{Recami.1976,IJMPA,IJMPB}
\begin{equation}
  [\hat{E}, \hat{t}] = i\hbar.
\label{eq.2.1.14}
\end{equation}

The Stone and von Neumann theorem\cite{Stone.1930}, %[29],
has been always interpreted as establishing a commutation relation like (\ref{eq.2.1.14}) for
the pair of the canonically conjugate operators (\ref{eq.2.1.1}) and (\ref{eq.2.1.13}), in
both representations, for selfadjoint operators only. However, it can be generalized for
(maximal) hermitian operators, once one introduces $\hat{t}$ by means of the {\em single-valued} Fourier transformation
from the $t$-axis ($-\infty < t < \infty$) to the $E$-semiaxis ($0 < E < \infty$), and utilizes the
properties\cite{Akhiezer} of the ``(maximal) hermitian" operators: This has been shown, e.g., in the last one of
Refs.\cite{Olkhovsky_Recami.1968.NuovoCim} as well as in Refs.\cite{IJMPA,IJMPB}.

Indeed, from eq.(\ref{eq.2.1.14}) the uncertainty relation
\begin{equation}
  \Delta E\; \Delta t \ge \hbar / 2
\label{eq.2.1.15}
\end{equation}
(where the standard deviations are $\Delta a = \sqrt{Da}$, quantity $Da$ being the variance
$Da = \langle a^{2} \rangle - \langle a \rangle^{2}$, \ and \  $a=E,t$, while
$\langle\ldots\rangle$ denotes the average over $t$ with the measures $W\,(x,t)\,dt$ or
$W_{\pm}\, (x,t)\,dt$ in the $t$-representation) can be derived also for operators which are simply hermitian, by a
straightforward generalization of the procedures which are common in the case of
{\em selfadjoint\/} (canonically conjugate) quantities, like coordinate $\hat{x}$ and momentum $\hat{p}_{x}$. Moreover, relation (\ref{eq.2.1.14})
satisfies\cite{IJMPA,IJMPB} the Dirac ``correspondence'' principle, since the classical Poisson brackets
$\{q_{0}, p_{0}\}$, with $q_{0}=t$ and $p_{0}=-E$, are equal to 1. \  In Refs.[4--7] and
\cite{IJMPA,IJMPB}, it was also shown that {\em the differences}, between the mean times at
which a wave-packet passes through a {\em pair} of points, obey the Ehrenfest correspondence
principle.

\

As a consequence, one can state that, for systems with continuous energy spectra, the
mathematical properties of (maximal) hermitian operators, like $\hat{t}$ in eq.(\ref{eq.2.1.1}),
are sufficient for considering them as quantum observables. Namely, the
{\em uniqueness\/}\cite{Akhiezer}  of the spectral decomposition (although not orthogonal)
for operators $\hat{t}$, and $\hat{t}^{n}$ ($n>1$), guarantees the ``equivalence" of
the mean values of any analytical function of time when evaluated in the $t$ and in the
$E$-representations. \ In other words, such an expansion is equivalent to a
completeness relation, for the (approximate) eigenfunctions of $\hat{t}^{n}$ ($n>1$),
which {\em with any accuracy} can be regarded as orthogonal, and corresponds to the
actual eigenvalues for the continuous spectrum. These approximate eigenfunctions belong
to the space of the square-integrable functions of the energy $E$ (cf., for instance,                                                                                                                          see, for instance
Refs.\cite{Recami.1976,Olkhovsky.1984.SJPN,Olkhovsky.1990.Nukleonika,IJMPA} and refs. therein).

From this point of view, there is no {\em practical} difference between selfadjoint and
maximal hermitian operators for systems with continuous energy spectra. Let us repeat that
the mathematical properties of $\hat{t}^{n}$ ($n>1$) are enough for considering time as
a quantum mechanical observable (like energy, momentum, space coordinates, etc.)
{\em without having to introduce any new physical postulates}.

\begin{figure}[!h]
\begin{center}
 \scalebox{2}{\includegraphics{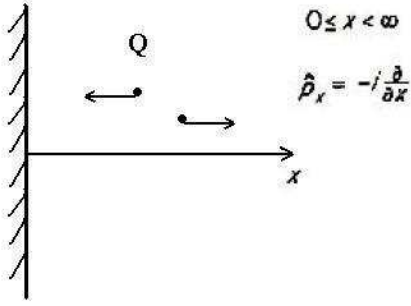}}
\end{center}
\caption{
For a particle Q free to move in a semi-space, bounded by a rigid wall located at $x=0$,
the operator $-i\partial / \partial x$ has the clear physical
meaning of  the particle momentum $x$-component even if it is {\em not} selfadjoint
(cf. von Neumann\cite{VonNeumann.1955}, and
Ref.\cite{Recami.1976}):  See the text.}
\label{fig.1}
\end{figure}

It is remarkable that von Neumann himself\cite{VonNeumann.1955}, % [32],
before confining himself for simplicity to selfajoint operators, stressed that operators like
our time $\hat{t}$ may
represent physical observables, even if they are not selfadjoint.  Namely, he explicitly considered the example of  the
operator $-\,i\hbar\, \partial / \partial x$ associated with a particle living in the right
semi-space bounded by a rigid wall located at $x=0$; that operator is not selfadjoint (acting
on wave packets defined on the
positive $x$-axis) only, nevertheless it obviously corresponds to the $x$-component of the observable {\em momentum} for
that particle: See Fig.{\ref{fig.1}}.

\

\

At this point, let us emphasize that our previously
assumed boundary condition $E \ne 0$ can be dispensed with, by
having recourse\cite{Olkhovsky_Recami.1968.NuovoCim,Recami.1976} to the
{\it bi-linear} hermitian operator
\begin{equation}
  \hat{t} =
  \displaystyle\frac{-i\hbar}{2}
  \displaystyle\frac{\stackrel{\leftrightarrow}{\partial}}{\partial E}
\label{eq.2.1.16}
\end{equation}
where the meaning of the sign $\leftrightarrow$ is clear from the accompanying definition
$$(f,\, \hat{t}\, g) = \Bigl(f,\: -\displaystyle\frac{ih}{2} \displaystyle\frac{\partial}{\partial E}\, g \Bigr) +
\Bigl( -\displaystyle\frac{ih}{2}\, \displaystyle\frac{\partial}{\partial E}\, f,\; g \Bigr) \, .$$
By adopting this expression for the time operator, the algebraic
sum of the two terms in the r.h.s. of the last relation results to be automatically zero at point
$E=0$. This question will be exploited below, in Sect.3 (when
dealing with the more general case of the four-position operator).
Incidentally, such an
``elimination"\cite{Recami.1976,Olkhovsky_Recami.1968.NuovoCim} of point
$E=0$ is not only simpler, but also more physical, than other
kinds of elimination obtained much later in papers like
\cite{Muga.1999}.

In connection with the last quotation, leu us for briefly comment on
the so-called {\em positive-operator-value-measure} (POVM)
approach, often used or discussed in the second set of papers on
time in quantum physics mentioned in our Introduction. Actually, an analogous procedure had been
proposed, since the sixties\cite{Aharonov.1961.PRA}, in some approaches to the quantum theory of measurements.
Afterwards, and much later, the POVM
approach has been applied, in a simplified and shortened form, to
the time-operator problem in the case of one-dimensional free
motion: for instance, in Refs.[10, 12, 15, 22--29] and especially
in~\cite{Muga.1999}. These papers stated that a generalized
decomposition of unity (or ``POV measure") could be obtained
from selfadjoint extensions of the time operator inside an extended
Hilbert space (for instance, adding the negative values of the energy, too), by exploiting the Naimark
dilation-theorem\cite{Naimark.1943}: But such a program has been realized till now
only in the simple cases of one-dimensional particle
free motion.

By contrast, our approach is based on a different
Naimark's theorem\cite{Naimark.1940}, which, as already mentioned
above, allows a much more direct, simple and general --and
at the same time non less rigorous-- introduction of a quantum
operator for Time. \ More precisely, our approach is based on the so-called {\em Carleman theorem\/}\cite{Carleman},
utilized in Ref.\cite{Naimark.1940}, about approximating a hermitian operator by suitable
successions of ``bounded" selfadjoint operators: That is, of selfadjoint operators whose
spectral functions do weakly converge to the non-orthogonal spectral function of the considered
hemitian operator. \ And our approach is applicable to a large family of three-dimensional (3D)
particle collisions, with all possible Hamiltonians. Actually, our approach was
proposed in the early Refs.[2--7] and in the first one of Ref.[18], and applied therein for
the time analysis of quantum collisions, nuclear reactions and tunnelling processes.

%-----------------------------------------------------------------------------------------------------------------------

%-----------------------------------------------------------------------------------------------------------------------
\subsection{On the momentum representation of the Time operator
\label{sec.2.2}}

In the continuous spectrum case, instead of the $E$-representation, with $0 < E <
+\infty$,  in eqs.(\ref{eq.2.1.1})--(\ref{eq.2.1.4}) one can also use the
$k$-representation\cite{Holevo.1978.RMP}, with the advantage that $-\infty < k < +\infty$:
\begin{equation}
  \Psi\, (x,t) =
  \displaystyle\int\limits_{-\infty}^{+\infty} g(k)\, \varphi(x,k)\, \exp(-iEt/\hbar)\: dk
\label{eq.2.2.1}
\end{equation}
with $E = \hbar^{2} k^{2} /\,2\mu$, and $k \ne 0$.

For the extension of the momentum representation to the case of $\langle t^{n} \rangle$,
with $n>1$, we confine ourselves here to refer the reader to the papers \cite{IJMPA,IJMPB}.
%-----------------------------------------------------------------------------------------------------------------------

%-----------------------------------------------------------------------------------------------------------------------
\subsection{An alternative weight for time averages (in the cases of particle {\em dwelling} inside
a certain spatial region)
\label{sec.2.3}}

We recall that the weight (\ref{eq.2.1.2}) [as well as its modifications (\ref{eq.2.1.3})]
has the meaning of a probability for the considered particle to pass through point $x$ during
the time interval ($t$, $t+dt$).  Let us follow the procedure
presented in Refs.[18-21],
%\cite{Olkhovsky_Recami.1992PR-1995JP-2004PR-2007IJMP,Olkhovsky.1997.conf,PhysRep2004,IJMPA},
 and refs. therein, and analyze the consequences of the equality
\begin{equation}
  \displaystyle\int\limits_{-\infty}^{+\infty}  j\,(x,t) \: dt =
  \displaystyle\int\limits_{-\infty}^{+\infty}  \bigl|\, \Psi (x,t) \bigr|^{2} \: dx
\label{eq.2.3.1}
\end{equation}
obtained from the 1D continuity equation. One can easily realize that a
second, alternative weight can be adopted:
\begin{equation}
  d\, P(x,t) \equiv
  Z\,(x,t)\: dx =
  \displaystyle\frac
    {\bigl| \Psi (x,t) \bigr|^{2} \: dx}
    {\displaystyle\int\limits_{-\infty}^{+\infty}  \bigl|\, \Psi (x,t) \bigr|^{2} \: dx}
\label{eq.2.3.2}
\end{equation}
which possesses the meaning of probability for the
particle to be located (or to sojourn, i.e., to
{\em dwell}) inside the infinitesimal space region ($x$, $x+dx$) at the
instant $t$, independently of its motion properties.  Then, the quantity
\begin{equation}
  P(x_{1},x_{2},t) =
  \displaystyle\frac
    {\displaystyle\int\limits_{x_{1}}^{x_{2}} \bigl| \Psi (x,t) \bigr|^{2} \: dx}
    {\displaystyle\int\limits_{-\infty}^{+\infty}  \bigl|\, \Psi (x,t) \bigr|^{2} \: dx}
\label{eq.2.3.3}
\end{equation}
will have the meaning of probability for the particle to dwell inside the spatial interval
($x_{1}$, $x_{2}$) at the instant $t$.

As it is known (see, for instance, Refs.\cite{Olkhovsky_Recami.1992PR-1995JP-2004PR-2007IJMP,PhysRep2004,IJMPA}
and refs. therein), the {\em mean dwell time} can be written in the {\em two}
equivalent forms:
\begin{equation}
  \langle \tau (x_{i}, x_{f}) \rangle =
  \displaystyle\frac
    {\displaystyle\int\limits_{-\infty}^{+\infty} dt \
     \displaystyle\int\limits_{x_{i}}^{x_{f}} |\Psi (x,t)|^{2} \; dx}
    {\displaystyle\int\limits_{-\infty}^{+\infty} j_{\rm in} (x_{i},t) \; dt}
\label{eq.2.3.4}
\end{equation}
and
\begin{equation}
  \langle \tau (x_{i}, x_{f}) \rangle =
  \displaystyle\frac
    {\displaystyle\int\limits_{-\infty}^{+\infty} t\, j(x_{f},t) \; dt -
     \displaystyle\int\limits_{-\infty}^{+\infty} t\, j(x_{i},t) \; dt}
    {\displaystyle\int\limits_{-\infty}^{+\infty} j_{\rm in} (x_{i},t) \; dt} \; ,
\label{eq.2.3.5}
\end{equation}
where it has been taken account, in particular, of relation
(\ref{eq.2.3.1}), which follows
---as already said--- from the continuity equation.

Thus, in correspondence with the two measures (\ref{eq.2.1.2}) and
(\ref{eq.2.3.2}), when integrating over time one gets {\em two}
different kinds of time distributions (mean values, variances,...),
which refer to the
particle traversal time in the case of measure (\ref{eq.2.1.2}),
and to the particle dwelling in the case of measure
(\ref{eq.2.3.2}). Some examples for 1D tunneling are contained in
Refs.\cite{Olkhovsky_Recami.1992PR-1995JP-2004PR-2007IJMP,PhysRep2004,IJMPA}.
%-----------------------------------------------------------------------------------------------------------------------

%-----------------------------------------------------------------------------------------------------------------------
\subsection{Time as a quantum-theoretical Observable in the case
of Photons
\label{sec.2.4}}

As is known (see, for instance, Refs.\cite{Schweber.1961,PhysRep2004}),
in first quantization the single-photon wave function can be
probabilistically described in the 1D case by the wave-packet\footnote{The gauge
condition ${\rm div} {\Abf} = 0$ is assumed.}
\begin{equation}
  {\Abf} ({\rbf}, t) =
    \displaystyle\int\limits_{k_{0}}
    \displaystyle\frac{d^{3}k}{k_{0}}\;
    {\chibf}({\kbf}) \: \varphi ({\kbf}, {\rbf})\: \exp(-ik_{0}t) \; ,
\label{eq.2.4.1}
\end{equation}
where, as usual, ${\Abf} ({\rbf},t)$ is the electromagnetic vector potential,
while ${\rbf}=\{x,y,z\}$, \
${\kbf} = \{k_{x},k_{y},k_{z}\}$, \ $k_{0} \equiv w/c = \varepsilon / \,\hbar c$, and
$k \equiv |{\kbf}| = k_{0}$. The axis $x$ has been chosen as the propagation direction. Let
us notice that ${\chibf} ({\kbf}) = \sum\limits_{i=y,z}
\chi_{i}({\kbf})\, {\ebf}_{i}({\kbf})$, \ with \ ${\ebf}_{i}
{\ebf}_{j} = \delta_{ij}$, and
$x_{i}, x_{j} = y,z$, while $\chi_{i}({\kbf})$ is the probability amplitude for the photon
to have momentum ${\kbf}$ and polarization ${\ebf}_{j}$ along $x_{j}$. Moreover, it is
$\varphi ({\kbf}, {\rbf}) = \exp(ik_{x}x)$ in the case of plane waves,
while $\varphi ({\kbf}, {\rbf})$ is a linear combination of evanescent (decreasing) and
anti-evanescent (increasing) waves in the case of ``photon barriers'' (i.e., band-gap filters,
or even undersized segments of waveguides for microwaves, or frustrated total-internal-reflection
regions for light, and so on). Although it is not easy to localize a photon in the direction
of its polarization\cite{Schweber.1961}, nevertheless for 1D propagations it is possible to
use the space-time probabilistic interpretation of eq.(\ref{eq.2.4.1}), and define the quantity
\begin{equation}
\begin{array}{cc}
  \rho_{\rm em} (x,t)\: dx = \displaystyle\frac{S_{0}\: dx} {\int  S_{0}\: dx}, &
  S_{0} = \displaystyle\int\displaystyle\int s_{0}\: dy\,dz
\end{array}
\label{eq.2.4.2}
\end{equation}
($s_{0} = [{\Ebf}^{*} \cdot {\Ebf} + {\Hbf}^{*} \cdot {\Hbf}]/\,4\pi$
being the  energy density, with the electromagnetic field  ${\Hbf}= {\rm rot}\,
{\Abf}$, \ and \ $\Ebf = -1/c \; \partial {\Abf} /  \partial t$), which represents
the probability density  {\em of a
photon to be found (localized) in the spatial interval ($x$, $x+dx$) along the $x$-axis
at the instant $t$}; \ and the quantity
\begin{equation}
\begin{array}{cc}
  j_{\rm em} (x,t)\: dt = \displaystyle\frac{S_{x}\: dt} {\int  S_{x}(x,t)\: dt}, &
  S_{x}(x,t) = \displaystyle\int\displaystyle\int s_{x}\: dy\,dz
\end{array}
\label{eq.2.4.3}
\end{equation}
($s_{x} = c\; \Re[{\Ebf}^{*} \wedge {\Hbf}]_{x}\, /
\:8\pi$ being the energy flux density), which represents {\em the
flux probability density of a photon to pass through point $x$
in the time interval ($t$, $t+dt$)}:  in full analogy with
the probabilistic quantities for non-relativistic
particles. The justification and convenience of such definitions
is self-evident, when the wave-packet group velocity coincides
with the velocity of the energy transport; \ in particular: \ (i)
the wave-packet (\ref{eq.2.4.1}) is quite similar to wave-packets
for non-relativistic particles, \ and \ (ii) in analogy with
conventional non-relativistic quantum mechanics, one can define
the ``mean time instant" for a photon (i.e., an electromagnetic
wave-packet) to pass through point $x$, as follows
\[
  \langle t(x) \rangle =
  \displaystyle\int\limits_{-\infty}^{+\infty} t\, J_{{\rm em},\, x}\; dt =
  \displaystyle\frac
    {\displaystyle\int\limits_{-\infty}^{+\infty} t\, S_{x} (x,t)\; dt}
    {\displaystyle\int\limits_{-\infty}^{+\infty} S_{x} (x,t)\; dt} \; .
\]
As a consequence [in the same way as in the case of equations (\ref{eq.2.1.1})--(\ref{eq.2.1.2})],
the form (\ref{eq.2.1.1}) for the time operator in the energy representation
is valid also for photons, with the same boundary conditions adopted in the case of particles, that is, with
$\chi_{i}\,(0) = \chi_{i}\,(\infty)$ and with $E= \hbar\, c\,k_{0}$.

The energy density $s_{0}$ and energy flux density $s_{x}$ satisfy the relevant continuity
equation
\begin{equation}
  \displaystyle\frac{\partial s_{0}}{\partial t} +
  \displaystyle\frac{\partial s_{x}}{\partial x} = 0
\label{eq.2.4.4}
\end{equation}
which is Lorentz-invariant for 1D spatial propagation\cite{PhysRep2004,IJMPA}
processes.
%-----------------------------------------------------------------------------------------------------------------------

%-----------------------------------------------------------------------------------------------------------------------
\subsection{Introducing the analogue of the ``Hamiltonian" for the case of the Time
operator: A new hamiltonian approach
\label{sec.2.5}}

In non-relativistic quantum theory, the Energy operator
acquires (cf., e.g., Refs.\cite{Olkhovsky.1990.Nukleonika,IJMPA}) the {\em two}
forms: \ (i) \ $i\hbar\, \displaystyle\frac{\partial}{\partial t}$
in the $t$-representation, and \ (ii) \ $\hat{H}\,(\hat{p}_{x},
\hat{x}, \ldots)$ in the hamiltonianian formalism. The ``duality" of these
two forms can be easily inferred from the Schr\"{o}edinger equation
itself,  \ $\hat{H}\Psi = i\hbar  \displaystyle\frac{\partial
\Psi}{\partial t}$. One can introduce in quantum mechanics a
similar duality for the case of {\em Time}: Besides  the general
form (\ref{eq.2.1.1}) for the Time operator in the energy
representation, which is valid for any physical systems in the
region of continuous energy spectra, one can {\em express the
time operator also in a ``hamiltonian form"}, i.e., in terms of the
coordinate and momentum operators, by having recourse to their
commutation relations. \ Thus, by the replacements
\begin{equation}
\begin{array}{c}
\vspace{3mm}
  \hat{E} \to \hat{H}\, (\hat{p}_{x}, \hat{x}, \ldots), \\
  \hat{t} \to \hat{T}\, (\hat{p}_{x}, \hat{x}, \ldots),
\end{array}
\label{eq.2.5.1}
\end{equation}
and on using the commutation relation [similar to eq.(\ref{eq.2.1.3})]

\begin{equation}
  [ \hat{H},\, \hat{T}] = i\hbar \; ,
\label{eq.2.5.2}
\end{equation}
one can obtain\cite{Rosenbaum.1969.JMP}, given a specific ordinary Hamiltonian, the corresponding
explicit expression for $\hat{T}\, (\hat{p}_{x}, \hat{x}, \ldots)$.

Indeed, this procedure can be adopted for any physical system with a known Hamiltonian
$\hat{H}\, (\hat{p}_{x}, \hat{x}, \ldots)$, and we are going to see a concrete example.
By going on from the coordinate to the momentum representation, one realizes that the
{\em formal} expressions of {\em both} the hamiltonian-type operators
$\hat{H}\,(\hat{p}_{x}, \hat{x}, \ldots)$ and
$\hat{T}\, (\hat{p}_{x}, \hat{x}, \ldots)$ {\em do not change}, except for an obvious change of sign
in the case of operator $\hat{T}\, (\hat{p}_{x}, \hat{x}, \ldots)$.

As an explicit example, let us address the simple case of a free particle whose Hamiltonian
is
\begin{equation}
  \hat{H} =
  \left\{
  \begin{array}{ccll}
    \hat{p}_{x}^{2}/\,2\mu, &
      \hat{p}_{x} =  -i\hbar  \displaystyle\frac{\partial}{\partial x} \, , &
      \mbox{ \ \ \ \ in the coordinate representation} & \mbox{(a)} \\
    p_{x}^{2}/\,2\mu \, . & &
      \mbox{ \ \ \ \ in the momentum representation}& \mbox{(a)}
  \end{array}
  \right.
\label{eq.2.5.3}
\end{equation}
Correspondingly, the Hamilton-type {\em time operator,} in its symmetrized form, will write
\begin{equation}
  \hat{T} =
  \left\{
  \begin{array}{clll}
    \vspace{2mm}
    \displaystyle\frac{\mu}{2}\,
    \Bigl( \hat{p}_{x}^{-1} x + x\hat{p}_{x}^{-1} + i\hbar \, ; \ \hat{p}_{x}^{-2} \Bigr), &
      {\rm in \ the \ coordinate \ representation} & \mbox{(a)} \\
    -\displaystyle\frac{\mu}{2}\,
    \Bigl( p_{x}^{-1} \hat{x} + \hat{x}p_{x}^{-1} + i\hbar / p_{x}^{2} \Bigr), &
      {\rm in \ the \ momentum \ representation} & \mbox{(b)}
  \end{array}
  \right.
\label{eq.2.5.4}
\end{equation}
where
\[
\begin{array}{cc}
  \hat{p}_{x}^{-1} = \displaystyle\frac{i}{\hbar} \int dx \ldots, &
\hspace{7mm}
  \hat{x} =  i\hbar \displaystyle\frac{\partial}{\partial p_{x}} \; .
\end{array}
\]
Incidentally, operator (\ref{eq.2.5.4}b) is equivalent to
$-i\hbar\, \frac{\partial}{\partial E}$,  since
$E=p_{x}^{2}/\,2\mu$;  and therefore it is also a (maximal)
{\em hermitian} operator. Indeed, by applying the operator $\hat{T}\,
(\hat{p}_{x}, \hat{x}, \ldots)$,  for instance, to a plane-wave of the type $\exp(ikx)$,
we obtain the same result in both
the coordinate and the momentum representations:
\begin{equation}
  \hat{T} \; \exp(ikx) = \displaystyle\frac{x}{v}\: \exp(ikx)
\label{eq.2.5.5}
\end{equation}
quantity ${x}/{v}$ being the free-motion time (for a particle with velocity $v$ )
for traveling the distance $x$.

On the basis of what precedes, it is possible to show
that the wave function $\Psi(x,t)$ of a quantum system satisfies the two (dual) equations
\begin{equation}
\begin{array}{ccc}
  \hat{H}\, \Psi = i\hbar \displaystyle\frac{\partial \Psi}{\partial t} &
  \mbox{and} &
  \hat{T}\, \Psi = t\, \Psi \, .
\end{array}
\label{eq.2.5.6}
\end{equation}
In the energy representation, and in the stationary case, we obtain again {\em two} (dual)
equations
\begin{equation}
\begin{array}{ccc}
  \hat{H}\, \varphi_{t} = \varepsilon\, \varphi_{t} &
  \mbox{and} &
  \hat{T}\, \varphi_{t} = -i\hbar \displaystyle\frac{\partial \varphi_{t}}{\partial \varepsilon} \; ,
\end{array}
\label{eq.2.5.7}
\end{equation}
quantity $\varphi_{t}$ being the Fourier-transform of $\Psi$:
\begin{equation}
  \varphi_{t} =
    \displaystyle\frac{1}{2\pi\hbar}
    \displaystyle\int\limits_{-\infty}^{+\infty}
    \Psi(x,t) \: e^{i\varepsilon t/ \hbar}\; dt \, .
\label{eq.2.5.8}
\end{equation}

It might be interesting to apply the two pairs of the last dual equations also for investigating
tunnelling processes through the quantum gravitational barrier, which appears during inflation,
or at the beginning of the big-bang expansion, whenever a quasi-linear Schroedinger-type equation
does approximately show up.
%-----------------------------------------------------------------------------------------------------------------------

%-----------------------------------------------------------------------------------------------------------------------
\subsection{Time as an Observable (and the Time-Energy uncertainty relation), for quantum-mechanical systems
with {\em discrete} energy spectra
\label{sec.2.6}}

For describing the time evolution  of non-relativistic quantum
systems endowed with a purely {\em discrete} (or a continuous {\em
and discrete\/}) spectrum, let us now introduce wave-packets of
the form\cite{Olkhovsky.1990.Nukleonika,IJMPA,IJMPB}:
\begin{equation}
  \psi\, (x,t) =
    \sum\limits_{n=0}
    g_{n}\, \varphi_{n}(x)\, \exp[-i(\varepsilon_{n} - \varepsilon_{0})t / \hbar] \; ,
\label{eq.2.6.1}
\end{equation}
where $\varphi_{n}(x)$ are orthogonal and normalized bound states
which satisfy the equation $\hat{H}\, \varphi_{n}(x) =
\varepsilon_{n}\, \varphi_{n}(x)$, quantity $\hat{H}$ being the
Hamiltonian of the system; while the coefficients $g_n$ are
normalized: $\sum\limits_{n=0} |g_{n}|^{2} = 1$. We omitted the
non-significant phase factor $\exp(-i \varepsilon_{0} t/\hbar)$
of the fundamental state.

\begin{figure}[!h]
\begin{center}
 \scalebox{2.0}{\includegraphics{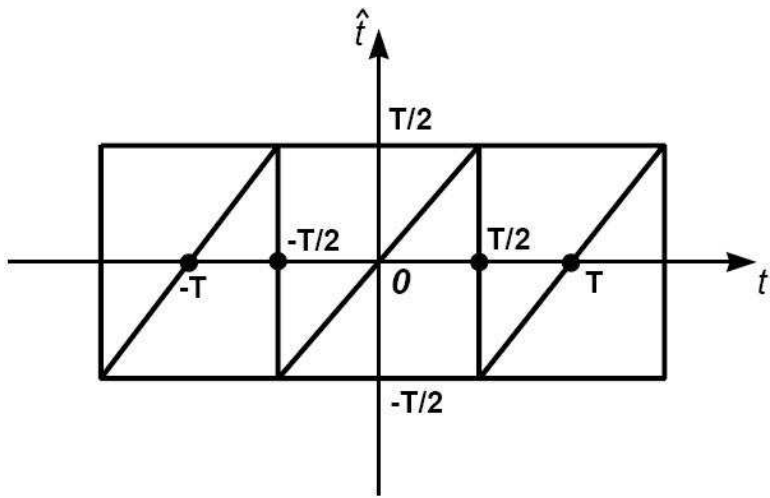}}
\end{center}
\caption{
The periodic saw-tooth function for the time operator in the case
of quantum mechanical systems with {\em discrete} energy spectra:
Namely, for the case of eq.(30).}
\label{fig.2}
\end{figure}

Let us first consider the systems whose energy levels are
separated by intervals admitting a maximum common divisor $D$ (for
ex., harmonic oscillator, particle in a rigid box, and
spherical spinning top), so that the wave packet
(\ref{eq.2.6.1}) is a periodic function of time possessing as
period the Poincar\'{e} cycle time $ T = 2\pi\hbar/D$. For such
systems it is possible\cite{Olkhovsky.1990.Nukleonika,IJMPA,IJMPB}
to construct a {\em selfadjoint} time operator with the form (in
the time representation) of a saw-function of $t$, choosing $t=0$
as the initial time instant:
\begin{equation}
\hat{t} \; = \; t - T \sum_{n=0}^\infty \Theta(t-[2n+1]T/2) +
T \sum_{n=0}^\infty \Theta(-t-[2n+1]T/2 \; .
\label{eq.2.6.1bis}
\end{equation}
This periodic function for the time operator is a linear (increasing) function of time $t$ within
each Poincarè cycle: see Fig.{\ref{fig.2}}.

\

\

The commutation relations of the Energy and Time operators, now both selfajoint, acquires in
the case of discrete energies and of a periodic Time operator the form
\begin{equation}
[\hat{E},\hat{t}]  \; = \; i \hbar \left\{ 1-T \sum_{n=0}^\infty \delta(t-[2n+1]T) \right\} \, ,
\label{eq.2.6.1ter}
\end{equation}
wherefrom the uncertainty relation follows in the new form
\begin{equation}
(\Delta E)^2 \; (\Delta t)^2 \; = \; \hbar^2 \left[ 1-{{T|\psi(T/2+\gamma)|^2}
\over {\int_{-T/2}^{T/2} |\psi(t)|^2 dt}} \right] \; ,
\label{eq.2.6.1quater}
\end{equation}
where it has been introduced a parameter $\gamma$, with $-T/2 < \gamma < T/2$, in order to
assure that the r.h.s. integral is single-valued\cite{IJMPA,IJMPB}.

When $\Delta E \rightarrow 0$ (that is, when $|g_n| \rightarrow
\delta_{n n'})$, the r.h.s. of eq.\ref{eq.2.6.1quater} tends to
zero too, since $|\psi(t)|^2$ tends to a constant value. In such a
case, the distribution of the time instants at which the
wave-packet passes through point $x$ becomes flat within each
Poincar\'e cycle. When, by contrast, $\Delta E >> D$ and \
$|\psi(T+\gamma)|^2 << (\int_{-T/2}^{T/2} |\psi(t)|^2 dt) / T$,
the periodicity condition may become inessential whenever $\Delta
t << t$. In other words, our uncertainty relation
(\ref{eq.2.6.1quater}) transforms into the ordinary uncertainty
relation for systems with continuous spectra.

In more general cases, for excited states of nuclei, atoms  and
molecules, the {\em energy-level intervals, for discrete and
quasi-discrete (resonance) spectra, are not multiples of a maximum
common divisor,} and hence the Poincar\'{e} cycle
is not well-defined for such systems. \ Nevertheless, even for those systems one can
introduce an approximate description (sometimes, with any desired
degree of accuracy) in terms of Poincar\'e quasi-cycles and a
quasi-periodical evolution; so that for sufficiently long time
intervals the behavior of the wave-packets can be associated with
a {\em a periodical motion (oscillation)}, sometimes ---e.g., for
very narrow resonances--- with any desired accuracy. For them,
when choosing an approximate Poincar\'{e}-cycle time, one can
include in one cycle as many quasi-cycles as it is necessary for
the demanded accuracy. Then, with the chosen accuracy, a {{\em
quasi-selfadjoint time operator}} can be introduced.
%-----------------------------------------------------------------------------------------------------------------------

%-----------------------------------------------------------------------------------------------------------------------
\section{On four-position operators in quantum field theory, in terms of
{\em bilinear} operators
\label{sec.3}}

In this Section we approach the {\em relativistic} case, taking into consideration
---therefore--- the space-time (four-dimensional) ``position" operator, starting
however with an analysis of the 3-dimensional (spatial) position operator in the
simple relativistic case of the Klein-Gordon equation.

Actually, this analyzis
will lead us to tackle already with non-hermitian operators. \ Moreover, while
performing it, we shall meet the opportunity of introducing bilinear operators,
which will be used even more in the next case of the full 4-position
operator.

Let us recall that in Sect.2.1 we mentioned that the boundary condition $E \ne 0$, therein
imposed to guarantee (maximal) hermitity of the time operator, can be dispensed with just by
having recourse to bilinear forms. \ Namely, by considering the bilinear hermitian
operator\cite{Recami.1976,Recami.1983.HJ}
 \ $\hat{t}=(-i\hbar \; {\stackrel{\leftrightarrow}{\partial}} /{\partial E})/2$, \
where the sign $\leftrightarrow$ is defined through the accompanying equality \
$(f,\, \hat{t}\, g) = \Bigl(f,\:
 -\frac{ih}{2}
\frac{\partial}{\partial E}\, g \Bigr) + \Bigl( -\frac{ih}{2}\, \frac{\partial}{\partial E}\,
f,\; g \Bigr)$.

\subsection{The Klein-Gordon case: Three-position operators
\label{sec.3.1}}

The standard position operators, being hermitian and moreover
selfadjoint, are known to possess real eigenvalues: i.e., they
yield a {{\em point-like}} localization. J.M.Jauch showed,
however, that a point-like localization would be in contrast with
``unimodularity". In the relativistic case, moreover, phenomena so
as the pair production forbid a localization with precision better
than one Compton wave-length. The eigenvalues of a realistic
position operator ${\hat{\zbf}}$ are therefore expected to
represent space {{\em regions}}, rather than points. This can be
obtained only by having recourse to non-hermitian (and therefore
non-selfadjoint) position operators ${\hat{\zbf}}$ (a priori, one
can have recourse either to non-normal operators with commuting
components, or to normal operators with non-commuting components).
Following, e.g., the ideas in Ref.\cite{Kalnay}, % [36],
we are going to show that the mean values of the {\em hermitian (selfadjoint) part} of
${\hat{\zbf}}$ will yield a mean (point-like) position\cite{Recami.1970}, % [37],
while the mean values of the {\em anti-hermitian (anti-selfadjoint) part} of ${\hat{\zbf}}$
will yield the sizes of the
localization region\cite{Olkhovsky_Recami.1968.NuovoCim}. % [2].

Let us consider, e.g., the case of relativistic spin-zero particles,
in natural units and with metric $(+\, -\, -\, -)$. The position
operator $i\, \nabla_{\imp}$, is known to be actually
non-hermitian, and may be in itself a good candidate for an
extended-type position operator. To show this, we want to
split\cite{Kalnay} it into its hermitian and anti-hermitian (or skew-hermitian) parts.

Consider, then, a vector space $V$ of complex differentiable
functions on a 3-dimensional phase-space\cite{Recami.1983.HJ} equipped with an inner
product defined by
\begin{equation}
  (\Psi,\, \Phi) =
    \displaystyle\int
    \displaystyle\frac{d^{3}{\imp}}{p_{0}}\;
    \Psi^{*}({\imp})\, \Phi(p)
\label{eq.3.1.1}
\end{equation}
quantity $p_{0}$ being $\sqrt{{\imp^{2}}+m_{0}^{2}}$. Let the functions in $V$ satisfy moreover the condition
\begin{equation}
  \lim \limits_{R \to \infty}
    \displaystyle\int\limits_{S_{R}}
    \displaystyle\frac{d S}{p_{0}}\;
    \Psi^{*}(p)\, \Phi(p) = 0
\label{eq.3.1.2}
\end{equation}
where the integral is taken over the surface of a sphere of radius
$R$. \ If $U : V \to V$ is a differential operator of degree one,
condition (\ref{eq.3.1.2}) allows a definition of the transpose
$U^{T}$ by
\begin{equation}
\begin{array}{ccc}
  (U^{T} \Psi,\, \Phi) =
  (\Psi,\, U\, \Phi) &
  \mbox{for all  } \Psi,\, \Phi \in V \; ,
\end{array}
\label{eq.3.1.3}
\end{equation}
where $U$ is changed into $U^{T}$, or vice-versa, by means of integration by parts.

This allows, further, to introduce a {\em dual representation\/}\cite{Recami.1983.HJ}
($U_{1}$, $U_{2}$) of a {{\em single}} operator $U_{1}^{T} +
U_{2}$ by
\begin{equation}
  (U_{1} \Psi,\, \Phi) + (\Psi,\, U_{2} \Phi) = (\Psi,\, (U_{1}^{T} + U_{2})\, \Phi).
\label{eq.3.1.4}
\end{equation}
With such a dual representation, it is easy to split any operator
into its hermitian and anti-hermitian parts
\begin{equation}
  (\Psi,\, U \Phi) =
  \displaystyle\frac{1}{2}\, \Bigl((\Psi,\, U \Phi) + (U^{*}\Psi,\, \Phi)\Bigr) +
  \displaystyle\frac{1}{2}\, \Bigl((\Psi,\, U \Phi) - (U^{*}\Psi,\, \Phi)\Bigr).
\label{eq.3.1.5}
\end{equation}
Here the pair
\begin{equation}
  \displaystyle\frac{1}{2} \, (U^{*}, \, U ) \equiv \, \stackrel{\leftrightarrow}{U}_{h} \; ,
\label{eq.3.1.6}
\end{equation}
corresponding to $(1/2)\, (U + U^{* T})$, represents the hermitian part, while
\begin{equation}
  \displaystyle\frac{1}{2} (- U^{*},\, U ) \equiv\, \stackrel{\leftrightarrow}{U}_{a}
\label{eq.3.1.7}
\end{equation}
represents the anti-hermitian part.

Let us apply what precedes to the case of the Klein-Gordon position-operator
$\hat{z} = i\, \nabla_{p}$. When
\begin{equation}
  U = i\, \displaystyle\frac{\partial}{\partial p_{j}}
\label{eq.3.1.8}
\end{equation}
we have\cite{Olkhovsky_Recami.1968.NuovoCim} % [2]
\begin{equation}
\begin{array}{cl}
\vspace{2mm}
  \displaystyle\frac{1}{2}\, (U^{*},\, U ) =
  \displaystyle\frac{1}{2}\,
    \biggl(
      -i\,\displaystyle\frac{\partial}{\partial p_{j}},\,
      i\,\displaystyle\frac{\partial}{\partial p_{j}}
    \biggr) \equiv\,
  \displaystyle\frac{i}{2}\,
    \displaystyle\frac{\stackrel{\leftrightarrow}{\partial}}{\partial p_{j}}, & (a) \\

  \displaystyle\frac{1}{2}\, (-U^{*},\, U ) =
  \displaystyle\frac{1}{2}\,
    \biggl(
      i\,\displaystyle\frac{\partial}{\partial p_{j}},\,
      i\,\displaystyle\frac{\partial}{\partial p_{j}}
    \biggr) \equiv\,
  \displaystyle\frac{i}{2}\,
    \displaystyle\frac{\stackrel{\leftrightarrow}{\partial}_{+}}{\partial p_{j}}. & (b)
\end{array}
\label{eq.3.1.9}
\end{equation}
And the corresponding {{\em single}} operators turn out to be
\begin{equation}
\begin{array}{ll}
\vspace{2mm}
  \displaystyle\frac{1}{2}\, (U + U^{*T}) =
  i\, \displaystyle\frac{\partial}{\partial p_{j}} -
  \displaystyle\frac{i}{2}\, \displaystyle\frac{p_{j}}{p^{2} + m_{0}^{2}}, & (a) \\

  \displaystyle\frac{1}{2}\, (U - U^{*T}) =
  \displaystyle\frac{i}{2}\, \displaystyle\frac{p_{j}}{p^{2} + m_{0}^{2}} \; . & (b)
\end{array}
\label{eq.3.1.10}
\end{equation}
It is noteworthy\cite{Olkhovsky_Recami.1968.NuovoCim} % [2]
that, as we are going to see, operator (\ref{eq.3.1.10}a) is nothing but the usual Newton-Wigner operator, while
(\ref{eq.3.1.10}b) can be
interpreted\cite{Kalnay,Olkhovsky_Recami.1968.NuovoCim,Toller.1999.PRA} % [21, 36]
as yielding the sizes of the localization-region (an ellipsoid) via
its average values over the considered wave-packet.

Let us underline that the previous formalism justifies from the mathematical point of view
the treatment presented in papers like~\cite{Kalnay,Recami.1970}. % [36, 37]
 \ We can split\cite{Olkhovsky_Recami.1968.NuovoCim}
the operator $\hat{z}$ into two {\em bilinear} parts, as follows:
\begin{equation}
  \hat{z} = i\, \nabla_{p} =
  \displaystyle\frac{i}{2}\, \stackrel{\leftrightarrow}{\nabla}_{p} +
  \displaystyle\frac{i}{2}\, \stackrel{\leftrightarrow}{\nabla}_{p}^{(+)}
\label{eq.3.1.11}
\end{equation}
where \ $\Psi^{*} \stackrel{\leftrightarrow}{\nabla}_{p} \Phi \equiv\,
\Psi^{*} \nabla_{p} \Phi - \Phi \nabla_{p} \Psi^{*}$ \ and \
$\Psi^{*} \stackrel{\leftrightarrow}{\nabla}_{p}^{(+)} \Phi \equiv\,
\Psi^{*} \nabla_{p} \Phi + \Phi \nabla_{p} \Psi^{*} \, ,$
and where we always referred to a suitable\cite{Kalnay,Recami.1970,Recami.1976,Recami.1983.HJ} space of
wave packets. % [36, 37].
Its hermitian part\cite{Kalnay,Recami.1970} % [36, 37]
\begin{equation}
 \hat{x} = \displaystyle\frac{i}{2}\, \stackrel{\leftrightarrow}{\nabla}_{p} \; ,
\label{eq.3.1.12}
\end{equation}
which was expected to yield an (ordinary) point-like localization, has been derived also
by writing explicitly
\begin{equation}
  (\Psi,\, \hat{x}\, \Phi) =
    i\, \displaystyle\int
    \displaystyle\frac{d^{3}p}{p_{0}}\;
    \Psi^{*}(p)\, \nabla_{p}\, \Phi(p)
\label{eq.3.1.13}
\end{equation}
and imposing hermiticity, i.e., imposing the reality of the diagonal elements. The calculations
yield
\begin{equation}
  \Re\, \bigl(\Phi,\, \hat{x}\, \Phi \bigr) =
    i\, \displaystyle\int
    \displaystyle\frac{d^{3}p}{p_{0}}\;
    \Phi^{*}(p)\, \stackrel{\leftrightarrow}{\nabla}_{p}\, \Phi(p) \; ,
\label{eq.3.1.14}
\end{equation}
suggesting to adopt just the Lorentz-invariant quantity (\ref{eq.3.1.12}) as a bilinear
hermitian position operator. \
Then, on integrating by parts (and due to the vanishing of the surface integral), we
verify that eq.(\ref{eq.3.1.12}) {\em is
equivalent to the ordinary Newton-Wigner operator:}

\begin{equation}
  \hat{x}_{h} \equiv\,
  \displaystyle\frac{i}{2}\, \stackrel{\leftrightarrow}{\nabla}_{p}\: \equiv\,
  i\, \nabla_{p} - \displaystyle\frac{i}{2}\, \displaystyle\frac{{\imp}}{p^{2} + m^{2}}
\; \equiv {\rm N-W} \, .
\label{eq.3.1.15}
\end{equation}
We are left with the (bilinear) anti-hermitian part
\begin{equation}
  \hat{y} = \displaystyle\frac{i}{2}\, \stackrel{\leftrightarrow}{\nabla}_{p}^{(+)}
\label{eq.3.1.16}
\end{equation}
whose {\em average values} over the considered state (wave-packet) can be regarded as
yielding\cite{Kalnay,Recami.1970,Recami.1976,Recami.1983.HJ}% [7, 9]
the sizes of an ellipsoidal localization-region.

After the digression associated with eqs.(\ref{eq.3.1.11})--(\ref{eq.3.1.16}), let us go back to
the present formalism, as expressed
by eqs.(\ref{eq.3.1.1})--(\ref{eq.3.1.10}).

In general, the extended-type position operator $\hat{z}$ will yeld
\begin{equation}
  \langle \Psi|\, \hat{z}\, |\Psi \rangle =
  (\alpha + \Delta\alpha) + i\, (\beta + \Delta \beta) \; ,
\label{eq.3.1.17}
\end{equation}
where $\Delta\alpha$ and  $\Delta \beta$
are the mean-errors encountered when measuring the point-like position and the sizes of the localization
region, respectively. It is interesting to evaluate the commutators ($i,j = 1,2,3$):
\begin{equation}
  \biggl(
    \displaystyle\frac{i}{2}\, \displaystyle\frac{\stackrel{\leftrightarrow}{\partial}}{\partial p_{i}},\,
    \displaystyle\frac{i}{2}\, \displaystyle\frac{\stackrel{\leftrightarrow}{\partial}_{(+)}}{\partial p_{j}}
  \biggr) =
  \displaystyle\frac{i}{2\,p_{0}^{2}}\,
    \biggl( \delta_{ij} - \displaystyle\frac{2\,p_{i}p_{j}}{p_{0}^{2}} \biggr) \, ,
\label{eq.3.1.18}
\end{equation}
wherefrom the noticeable ``uncertainty correlations'' follow:
\begin{equation}
  \Delta\alpha_{i} \ \Delta\beta_{j} \ge
  \displaystyle\frac{1}{4} \;
  \biggl| \biggl\langle
    \displaystyle\frac{1}{p_{0}^{2}}\;
    \biggl( \delta_{ij} - \displaystyle\frac{2\,p_{i}p_{j}}{p_{0}^{2}} \biggr)
  \biggr\rangle \biggr| \, .
\label{eq.3.1.19}
\end{equation}
%-----------------------------------------------------------------------------------------------------------------------

%-----------------------------------------------------------------------------------------------------------------------
\subsection{Four-position operators
\label{sec.3.2}}

It is tempting to propose as {\em four-position operator} the quantity $\hat{z}^{\mu} = \hat{x}^{\mu} +
i\,\hat{y}^{\mu}$, whose hermitian (Lorentz-covariant) part can be written
\begin{equation}
  \hat{x}^{\mu} =
  -\,\displaystyle\frac{i}{2}\,
  \displaystyle\frac{\stackrel{\leftrightarrow}{\partial}}{\partial p_{\mu}} \; ,
\label{eq.3.2.1}
\end{equation}
to be associated with its corresponding ``operator'' in four-momentum space
\begin{equation}
  \hat{p}_{h}^{\mu} =
  +\: \displaystyle\frac{i}{2}\,
  \displaystyle\frac{\stackrel{\leftrightarrow}{\partial}}{\partial x_{\mu}} \; .
\label{eq.3.2.2}
\end{equation}

Let us recall the proportionality between the 4-momentum operator and the 4-current density
operator in the chronotopical space, and then underline the canonical correspondence (in the
4-position and 4-momentum spaces, respectively) between the ``operators" (cf. the previous subsection):
\begin{equation}
\begin{array}{ll}
\vspace{2mm}
  m_{0}\, \hat{\rho}\, \equiv\, \hat{p}_{0} =
  \displaystyle\frac{i}{2}\,
  \displaystyle\frac{\stackrel{\leftrightarrow}{\partial}}{\partial t}  & (a) \\

  m_{0}\, \hat{{\jbf}}\, \equiv\, \hat{{\imp}} =
  - \displaystyle\frac{i}{2}\,
  \displaystyle\frac{\stackrel{\leftrightarrow}{\partial}}{\partial {\rbf}} \; , & (b)
\end{array}
\label{eq.3.2.3}
\end{equation}
and the operators
\begin{equation}
\begin{array}{ll}
\vspace{2mm}
  \hat{t}\, \equiv\,
  - \displaystyle\frac{i}{2}\,
  \displaystyle\frac{\stackrel{\leftrightarrow}{\partial}}{\partial p_{0}} & (a) \\

  \hat{{\xbf}}\, \equiv\,
  \displaystyle\frac{i}{2}\,
  \displaystyle\frac{\stackrel{\leftrightarrow}{\partial}}{\partial {\imp}} \; , & (b)
\end{array}
\label{eq.3.2.4}
\end{equation}
where the four-position ``operator" (\ref{eq.3.2.4}) can be considered as a 4-current
density operator in the energy-impulse space. \ Analogous considerations can be carried on
for the anti-hermitial parts (see the last one of Refs.\cite{Olkhovsky_Recami.1968.NuovoCim}).

Finally, by recalling the properties of the time operator as a maximal hermitian operator in the
non-relativistic case (Sec.\ref{sec.2.1}), one can see that the relativistic
time operator (\ref{eq.3.2.4}a) (for
the Klein-Gordon case) is also a selfadjoint bilinear operator for the case of continuous
energy spectra, and a
(maximal) hermitian linear operator for free particles [due to the presence of the lower limit zero
for the kinetic energy, or $m_{0}$ for the total energy].

% *******************************************************************************************************************

% *******************************************************************************************************************
% \newpage
\section{Unstable states and non-hermitian Hamiltonians
\label{sec.4}}

\subsection{Introduction
\label{sec.4.1}}

This whole Section is based on work performed in
collaboration with A.Agodi, M.Baldo, and A.Pennisi di
Floristella.\cite{Agodi_Baldo_Recami.1973.AP}

In quantum mechanics the ``Resonance'' peaks are generally described as corresponding to unstable states (remember, e.g.,
Schwinger's approach\cite{Schwinger.1960.AP}). %[2]).
The present attempt proceeds as follows: (i) singling out {\em one} state $|\varphi\rangle$ in the state space; (ii)
finding out the effect of the (internal, virtual) state $|\varphi\rangle$ on the transition-amplitude; (iii) finding, in
particular, the necessary conditions for $|\varphi\rangle$ to be connected with a Resonance in the cross-section. In this
way one can associate the ``resonant state'' with the eigenvectors of a non-hermitian Hamiltonian (or rather, for
simplicity, of a  ``{\em quasi} hermitian'' Hamiltonian), such eigenvectors decaying correctly in time. We shall adopt the
formalism introduced by Akhiezer and Glazman\cite{Akhiezer}, %[3],
by Lifshitz, by Galitsky and Migdal\cite{Migdal.1958.JETP}, %[4],
and by Agodi et al.\cite{Agodi}. % [5].

Chosen a state $|\,\varphi\rangle$, let us define the projectors
\begin{equation}
\begin{array}{ll}
  P \equiv |\,\varphi\rangle\; \langle\varphi|, &
  Q = 1 - P \; ,
\end{array}
\label{eq.5.1.1}
\end{equation}
where 1 is the identity operator.
%-----------------------------------------------------------------------------------------------------------------------

%-----------------------------------------------------------------------------------------------------------------------
\subsection{Preliminary case: time-dependent description of potential scattering
\label{sec.4.2}}

Let us preliminarily consider the time-dependent description of potential scattering. Quantity $V$ be the potential
operator. In the limiting case of plane-waves, the scattering amplitude writes
\begin{equation}
\begin{array}{ll}
  T({\kbf}, {\kbf}') =
  \langle {\kbf}'|\, V\, |\,{\kbf} \rangle +
  \langle {\kbf}'|\, V\, G(E^{+})\,V\, |\,{\kbf} \rangle
\end{array}
\label{eq.5.2.1}
\end{equation}
with
\begin{equation}
\begin{array}{ll}
  G(E^{+}) \equiv (E^{+} - H)^{-1}, &
  E^{\pm} \equiv E \pm i\varepsilon  \; .
\end{array}
\label{eq.5.2.2}
\end{equation}
Chosen the {\em exploring} vector $|\,\varphi\rangle$ and using definition (\ref{eq.5.1.1}), we have
\begin{equation}
\begin{array}{ll}
  H = \stackrel{0}{H} + \stackrel{1}{H}, \\
  \stackrel{0}{H} = QHQ, & \stackrel{1}{H} = PHP + PHQ + QHP \; .
\end{array}
\label{eq.5.2.3}
\end{equation}
By introducing the scattering states $|\stackrel{0}{\varphi} \rangle$ due to $\stackrel{0}{H}$
\begin{equation}
  |\,\varphi^{\pm}_{{\kbf}} \rangle =
  \Biggl\{ 1 + \displaystyle\frac{1}{E^{\pm} - \stackrel{0}{H}}\,
    \Bigl(\stackrel{0}{H} - E\Bigr) \Biggr\} \ |\,{\kbf}\rangle,
\label{eq.5.2.4}
\end{equation}
we obtain
\begin{equation}
\begin{array}{lcl}
  S ({\kbf},{\kbf}') & \equiv &
  \langle \varphi_{{\kbf}'}^{(-)} |\,\varphi_{{\kbf}}^{(+)} \rangle =
  \langle \stackrel{0}{\varphi_{{\kbf}'}}{}^{(-)} |\,\stackrel{0}{\varphi_{{\kbf}}}{}^{(+)} \rangle -
  2\pi\, \delta(E_{{\kbf}'} - E_{{\kbf}}) \\

  & \times & \langle \stackrel{0}{\varphi_{{\kbf}'}}{}^{(-)} |\: HP\,G(E^{+}_{{\kbf}})\,PH\:
    |\stackrel{0}{\varphi_{{\kbf}}}{}^{(+)} \rangle \; ,
\end{array}
\label{eq.5.2.5}
\end{equation}

\noindent
where the first addendum in the r.h.s. of eq.(\ref{eq.5.2.5}) (let us call it $A$) is the contribution coming from
processes developing entirely in the subspace onto which $Q$ projects, whilst the second addendum ($B$) is contributed by
processes going through the exploring state $|\,\varphi\rangle$ onto which $P$ projects. In other words, the processes
with $|\,\varphi\rangle$ as intermediate state correspond to the term
\begin{equation}
\begin{array}{ll}
  \Bigl(2\pi\, \delta(E_{{\kbf}'} - E_{{\kbf}})\Bigr)^{-1} \ B =
  -2\pi i \;
  \displaystyle\frac
    {\langle \stackrel{0}{\varphi}{\hspace{-1mm}}_{{\kbf}'}^{(-)} |\, H\, |\,\varphi\rangle\;
    \langle \varphi|\,H\, |\stackrel{0}{\varphi}{\hspace{-1mm}}_{{\kbf}}^{(+)}\rangle}
    {E_{{\kbf}}^{+} - \langle \varphi |\, H\, |\, \varphi\rangle -
      \langle \varphi|\, W^{\varphi}(E_{{\kbf}}^{+})\, |\, \varphi\rangle}, \\
  W^{\varphi}(z) = PHQ\: \displaystyle\frac{1}{z - QHQ}\: QHP \; .
\end{array}
\label{eq.5.2.6}
\end{equation}
Our problem is: under what conditions one (or more) Resonances are actually associated with the chosen
$|\,\varphi\rangle$?  Let us notice in particular that, if
$E_{\varphi} \equiv
\langle\varphi|\,H\,|\,\varphi\rangle - \Re\, \langle\varphi|\, W^{\varphi}(E^{+})\, |\,\varphi\rangle$
and
$\Gamma_{\varphi} \equiv \Im\, \langle\varphi|\, W^{\varphi}(E^{+})\, |\,\varphi\rangle$
are smooth functions of $E$, then $B$ gets just the ``Breit and Wigner'' form:
\begin{equation}
  B =
  -2\pi i \;  \displaystyle\frac
    {\langle \stackrel{0}{\varphi}{\hspace{-1mm}}_{{\kbf}'}^{(-)} |\, HPH\,
    |\stackrel{0}{\varphi}{\hspace{-1mm}}_{{\kbf}}^{(+)}\rangle}
    {E - E_{\varphi} + i\,\Gamma_{\varphi}} \; .
\label{eq.5.2.7}
\end{equation}
%-----------------------------------------------------------------------------------------------------------------------

%-----------------------------------------------------------------------------------------------------------------------
\subsection{Case of central potential and spin-free particles
\label{sec.4.3}}

Let us choose the angular-momentum representation. If $|\,\varphi\rangle$ is assumed to be in particular invariant under
O(3), then both terms in which $S$ was split are diagonal. If $\stackrel{0}{\delta_{2}}$ are the phase-shifts due to
$QHQ$ and $\mu$ is the reduced mass, then
\begin{equation}
  S_{2}({\kbf}) \equiv
  \exp(2i \,\delta_{2}({{\kbf}})) = F_{2}({\kbf}) \;
  \exp(2i \stackrel{0}{\delta_{2}}({\kbf})) \; ,
\label{eq.5.3.1}
\end{equation}
with
\begin{equation}
  F_{2}({\kbf}) \equiv
  1 -
  \displaystyle\frac{2\pi i \mu}{2{\kbf}} \;
  \displaystyle\frac
    {\Bigl| \langle \varphi |\, H\, |\stackrel{0}{\varphi}{\hspace{-1mm}}^{(+)}_{klm}\rangle \Bigr|^{2}}
    {E^{+} - \langle \varphi |\, H\, |\, \varphi\rangle -
      \langle \varphi |\, W^{\varphi}(E_{{\kbf}}^{+})\, |\, \varphi\rangle} \; .
\label{eq.5.3.2}
\end{equation}
Let us observe that the phase-shift of $F_{2}({\kbf})$ crosses the value $\frac{1}{2}\, \pi$ (with positive slope)
when
\begin{equation}
  F_{2}({\kbf}) = -1.
\label{eq.5.3.3}
\end{equation}
The condition for a Resonance to appear are particularly transparent for $l=0$:
\begin{equation}
  F_{0}({\kbf}) =
  \displaystyle\frac
    {E - E_{\varphi} - i\,\lambda_{0}({\kbf})}
    {E - E_{\varphi} + i\,\lambda_{0}({\kbf})} \; ,
\label{eq.5.3.4}
\end{equation}
when
\begin{equation}
  \lambda_{0}({\kbf}) \equiv
  - \Im\, \langle\varphi |\, W^{\varphi}(E_{{\kbf}}^{+})\, |\, \varphi\rangle =
    \Bigl| \langle\varphi |\, H\, | \stackrel{0}{\varphi}{\hspace{-1mm}}_{k00}^{(+)} \rangle\Bigr|^{2}
\label{eq.5.3.5}
\end{equation}
is positive-definite. Namely, the condition $F_{0}({\kbf}) =-1$ yields
\begin{equation}
  |1 - S_{0}({\kbf})|^{2} = 4\, \cos^{2} \stackrel{0}{\delta}_{0} \; ,
\label{eq.5.3.6}
\end{equation}
with the supplementary conditions $\lambda_{0}({\kbf}) \ne 0$ and  $\cos \stackrel{0}{\delta}_{0} \ne 0$. When $\cos
\stackrel{0}{\delta}_{0} = 1$ the scattering due to $QHQ$ is negligible, i.e., the scattering proceeds entirely via the
intermediate formation of the (quasi-bound) state $|\,\varphi\rangle$; and the possible resonant effects are really
related to $|\,\varphi\rangle$. Of course, $\cos \stackrel{0}{\delta}_{0} = 1$ when, {\em at the resonance}
\begin{equation}
\begin{array}{ll}
  E=E_{\varphi}, & F({\kbf}) = -1 \; ,,
\end{array}
\label{eq.5.3.7}
\end{equation}
it is $|\stackrel{0}{\varphi}{\hspace{-1mm}}_{klm}^{(\pm)} \rangle = |\,lm \rangle$.

Notice that with every fixed $|\,\varphi\rangle$ a {\em series} of Resonances (also for different value of $l$) may be a
priori associated, if they are not destroyed by the $\stackrel{0}{\delta}_{0}$ behavior.
%-----------------------------------------------------------------------------------------------------------------------

%-----------------------------------------------------------------------------------------------------------------------
\subsection{Resonance definition
\label{sec.4.4}}

It is essential to recognize that the ``resonance condition'' $F_{l}({\kbf})=-1$ may be written
\cite{Agodi_Baldo_Recami.1973.AP}% [1]
\begin{equation}
  1 - \alpha\, ({\kbf},l)\: \langle\varphi_{2}|\, G (E^{+})\, |\,\varphi_{2} \rangle = 0 \; ,
\label{eq.5.4.1}
\end{equation}
with
\begin{equation}
  \alpha\, ({\kbf},l) =
  \displaystyle\frac{i\pi\mu}{2{\kbf}}\;
  \Bigl|\langle\varphi_{2} |\, H\, | \stackrel{0}{\varphi}{\hspace{-1mm}}^{(+)}_{klm} \rangle\Bigr|^{2}.
\label{eq.5.4.2}
\end{equation}

Let us now study the more general equation
\begin{equation}
  1 - \lambda\; \langle\varphi_{2}|\, G(z)\, |\,\varphi_{2}\rangle = 0 \; ,
\label{eq.5.4.3}
\end{equation}
where $z$, $\lambda$ are complex numbers.
Of course, a Resonance will appear at $\Re\, (z)$ if $z$ is near the real axis and if
\[
  \lambda = \alpha\,({\kbf},l) \; ,
\]
both satisfying eq.(\ref{eq.5.4.3}).

If we introduce at this point the non-hermitian (``quasi-hermitian") hamiltonian-operator
\begin{equation}
  H \equiv H + \lambda\, P \; ,
\label{eq.5.4.4}
\end{equation}
where $\lambda$ is complex and the ``resolvent operator'' is
\begin{equation}
  G(z) = \displaystyle\frac{1}{z - H} \; ,
\label{eq.5.4.5}
\end{equation}
then eq.(\ref{eq.5.4.3}) becomes
\begin{equation}
  \displaystyle\frac
    {\langle\varphi_{2}|\, G(z)\, |\,\varphi_{2} \rangle}
    {\langle\varphi_{2}|\, G(z)\, |\,\varphi_{2} \rangle} = 0 \; .
\label{eq.5.4.6}
\end{equation}
In other words, studying the (necessary) conditions for Resonance appearance is just equivalent to find out the poles in
the diagonal elements of the ``resolvent'' $G$-matrix, i.e., the eigenvalues of the {\em quasi selfadjoint} operator
$H$. \ Notice that, since
\begin{equation}
  G = G +
  G\: \displaystyle\frac
    {\lambda\, P}
    {1 - \lambda\, \langle\varphi_{2} |\, G(z)\, |\,\varphi_{2} \rangle}\: G \; ,
\label{eq.5.4.7}
\end{equation}
the difference between the spectra of $H$ and $H$ is just the presence of complex eigenvalues (corresponding to the
solution of our ``condition'' (\ref{eq.5.4.6})).

Therefore, in our framework the ``resonant (decaying) state'' $|\,\varphi\rangle$ is expected to be an eigenvector of $H$
(notice that it does {not} coincide with the state $|\,\varphi\rangle$ which is not unstable), corresponding to
the complex energy~$\cal E$.
%-----------------------------------------------------------------------------------------------------------------------

%-----------------------------------------------------------------------------------------------------------------------
\subsection{Applications
\label{sec.4.5}}

Let us confine ourselves to the case $l=0$, and rewrite the non-hermitian (quasi selfadjoint) hamiltonian as
\begin{equation}
\begin{array}{ll}
  H \equiv H + i\,\alpha_{{\kbf}}\, |\,\varphi\rangle\, \langle\varphi|, &
  \alpha_{{\kbf}} = -i\,\alpha\,({\kbf},0)
\end{array}
\label{eq.5.5.1}
\end{equation}
where
\begin{equation}
  V_{\varphi} = i\,\alpha_{{\kbf}}\, |\,\varphi\rangle\, \langle\varphi|
\label{eq.5.5.2}
\end{equation}
is anti-hermitian. We shall therefore write
\begin{equation}
  (H-{\cal E})\, |\,\psi\rangle =
  -V_{\varphi}\, |\,\psi\rangle \equiv
  - |\,\varphi\rangle \ i\,\alpha_{{\kbf}}\, \langle\varphi|\,\psi\rangle \; ,
\label{eq.5.5.3}
\end{equation}
which immediately yields for the eigenvalues the ``dispersion-type relation'' (${\cal E} \equiv {\cal E}_{\varphi}$):
\begin{equation}
  1 + i\, \biggl\langle\varphi\biggl|\, \displaystyle\frac{1}{H-{\cal E}}\,
    \biggl|\,\varphi\biggl\rangle\, \alpha_{{\kbf}} = 0 \; ,
\label{eq.5.5.4}
\end{equation}
and for the eigenvectors the explicit expression
\begin{equation}
  |\,\psi\rangle =
  - \langle\varphi |\, \psi\rangle\: i\alpha_{{\kbf}} \displaystyle\frac{1}{H-{\cal E}}\: |\,\varphi\rangle \; ,
\label{eq.5.5.5}
\end{equation}
where $\langle \varphi|\,\psi \rangle$ is a normalization constant. Notice that to solve eq.(\ref{eq.5.5.4}) we do not
need knowing $\alpha_{{\kbf}}$, i.e. the scattering states due to $QHQ$, since fortunately {\em at the resonances}
it is ($E=E_{R}$):
\[
\begin{array}{lll}
  \alpha_{{\kbf}} =
  \Bigl|\langle \varphi|\, H\, |\stackrel{0}{\psi}{\hspace{-1mm}}^{(+)}_{k00} \rangle \Bigr|^{2}=
  \Bigl|\langle \varphi|\,\psi^{(+)}_{E} \rangle -
    \langle \varphi|\,k00 \rangle \Bigr|^{-2}.
\end{array}
\]
Notice moreover that the present approach, a priori, allows distinguishing between true resonances and other effects.

In Ref.\cite{Agodi_Baldo_Recami.1973.AP} % [1]
the application was considered to the case of scattering by a
spherical-well potential $U(r) = U_{0} \; \theta(a-r)$, and as
exploring states the class was adopted of the normalized
Laurentian wave-packets (good for low energies)
\[
\begin{array}{lll}
  \langle k00|\, \varphi \rangle = \sqrt{2b}\, \displaystyle\frac{1}{k^{2}+b^{2}} &
  \Leftrightarrow &
  \langle r|\, \varphi \rangle = \sqrt{\displaystyle\frac{b}{2\pi}}\, \displaystyle\frac{\exp{(-br)}}{r} \; .
\end{array}
\]
By integration, for low energies ($k^{2} \ll 2mU_{0}$) one gets one equation whose real and imaginary parts forward a
system of two equations. The latter individuate $|\,\varphi\rangle$, i.e. the parameter $b$, for which a series of
(true) Resonances arises. These Resonances are expected to appear for ($k^{2} = 2mE$, $K^{2} = 2m\,(E+U_{0})$)
\[
\begin{array}{lll}
  \cos Ka = 0 & \to &
  K = \biggl(n+\displaystyle\frac{1}{2} \biggr)\, \pi \; .
\end{array}
\]
This system of equations is rather complicated (even when the resonance width is $\gamma < k_{0}$). But the first equation
does not contain $\gamma$ and yields $b$. For instance, for $n=0$ one gets a single solution ($ab \simeq 0.69$).
%-----------------------------------------------------------------------------------------------------------------------

%-----------------------------------------------------------------------------------------------------------------------
\subsection{Decay of the unstable state
\label{sec.4.6}}

We are more interested in the decay in time of the unstable state $|\,\psi\rangle$:
\begin{equation}
  \langle\psi|\,\psi_{t} \rangle \equiv
  \langle\psi|\, U_{t} \,|\,\psi \rangle \equiv
  \langle\psi|\, \exp{(-i{O}t)} \,|\,\psi \rangle.
\label{eq.5.6.1}
\end{equation}
If we assume, as usual, $O=H$, then
\begin{equation}
  \langle\psi|\,\psi_{t} \rangle =
  \displaystyle\int\limits_{0}^{\infty}
    \Bigl| \langle\psi|\, \psi_{E}^{(+)}\rangle\Bigr|^{2} \; \exp{(-i\,Et)}\; dE
\label{eq.5.6.2}
\end{equation}
since the bound-states do not contribute for large $t$. Moreover, let us remember that
\[
  |\,\psi\rangle =
  - i\alpha_{{\kbf}}\, \langle\varphi|\, \psi\rangle\: \displaystyle\frac{1}{H-{\cal E}}\: |\,\varphi\rangle \; .
\]
Therefore:
\[
\begin{array}{ll}
  \Bigl| \langle\psi_{E}^{(+)} |\, \psi\rangle\Bigr|^{2} =
    C \; \displaystyle\frac{|\alpha_{{\kbf}}|^{2}} {(\Re\, ({\cal E}) - E)^{2} -
    (\Im\, ({\cal E}))^{2}}, &
   \ \ \ C = \Bigl| \langle\psi_{E}^{(+)} |\, P |\,\psi\rangle\Bigr|^{2} \; .
\end{array}
\]
The integral (\ref{eq.5.6.2}) can be evaluated following Ref.\cite{Migdal.1958.JETP}. % [4].
The expression $C$ contains denominators that --- analytically extended --- produce {\em one} pole in $E={\cal E}$. If
in the strip $\Im\,({\cal E}) < \Im\,(E) <0$ no other singularities arise from the remaining factors, then we obtain the
exponential-type decay
\begin{equation}
  \langle\psi|\,\psi_{t} \rangle =
  \bigl(C + Dt\bigr) \; \exp \bigl(-iE_{0}t - \gamma_{0} t \bigr)
\label{eq.5.6.3}
\end{equation}
with $E_{0}= \Re\,({\cal E})$, \ $\gamma_{0}=\Im\,({\cal E})$, \ $C$ and $D$ {\em constants}.

More interesting appears, however, the assumption
\begin{equation}
  {O} = H
\label{eq.5.6.4}
\end{equation}
since in this case our approach does surely possess a ``Lie-admissible''
structure\cite{Santilli.1979.HJ} % [6]
(due to the fact that the time-evolution operator with $H$ is not unitary). In such a case one would simply get
\begin{equation}
  \langle\psi|\,\psi_{t} \rangle =
  \bar{K} \; \exp \bigl(iE_{0}t + \gamma_{0} t \bigr)
\label{eq.5.6.5}
\end{equation}
with $\bar{K} = \langle\psi|\, \psi\rangle$. But in this case the whole approach ought to be carefully rephrased in
Lie-admissible terms (otherwise, e.g. {\em all} states would seem to be decaying).

For instance, following Ref.\cite{Santilli.1979.HJ} % [6]
and with the choice (\ref{eq.5.6.4}), we may look for a hermitian operator $\tilde{H}$ such that the operator evolution
law becomes
\[
  A(t) = e^{itH}\, A(0)\, e^{-itH^{+}} =
    e^{it\tilde{H}S}\, A(0)\, e^{-itR\tilde{H}}
\]
where $\tilde{H}^{+}=\tilde{H}$, \ $S^{+}=R$. In particular, let us identify $\tilde{H}$ with $H$. One can verify, first of
all, that it does {not} seem possible to reduce ourselves to the Lie-isotopic
case\cite{Santilli.1979.HJ},
% [6],
in which $R=S \equiv g$ \ with \ $g^{+}=g$. In fact, in that case the operator $g$ ought to be
\[
  g = 1 + i\, \alpha H^{-1}\, |\,\varphi\rangle \langle\varphi| =
      1 - i\, \alpha\, |\,\varphi\rangle \langle\varphi| H^{-1}
\]
and this would imply
$\langle\varphi|\, H\, |\,\varphi\rangle = \langle\varphi|\, H^{-1}\, |\,\varphi\rangle = 0$.

In the more general (``Lie-admissible'') case, however, the relations ($S^{+} \ne S$)
\[
\begin{array}{ll}
  H     = H + i\, \alpha\, |\,\varphi\rangle \langle\varphi| = H S, &
  H^{+} = H - i\, \alpha\, |\,\varphi\rangle \langle\varphi| = S^{+} H
\end{array}
\]
allow immediately to write
\[
  R = S^{+} = 1 - i\, \alpha\, |\,\varphi\rangle\, \langle\varphi|\, H^{-1}.
\]
% *******************************************************************************************************************

% *******************************************************************************************************************
% \newpage
\section{Further examples of non-hermitian Hamiltonians: The cases of the nuclear optical
model, and of microscopic quantum dissipation
\label{sec.5}}

\subsection{Nuclear optical model
\label{sec.5.1}}

Since the fifties, the so-called optical model has been frequently used for describing
the experimental data on nucleon-nucleus elastic scattering, and, not less, on more general
nuclear collisions: see, e.g., Refs.\cite{Feshbach,Hodgson,Moldauer,Koning}; while for a
generalized optical model ---namely, the coupled-channel method with an optical model
in any channel of the nucleon-nucleus (elastic or inelastic) scattering, one can see Ref.\cite{Kunieda} and
refs. therein.

In the previous cases, the Hamiltonian contains a complex potential, its imaginary part
describing the absorption processes that take place by compound-nucleus formation and
subsequent decay. As to the Hamiltonian with complex potential, here we confine
ourselves at referring to work of ours already published, where it was studied the
non-unitarity and analytical structure of the $S$-matrix, the completeness of the
wave-functions, and so on: see Ref.\cite{Nikolayev}, and also
\cite{Olkhovsky1,Olkhovsky2}.

\subsection{Microscopic quantum dissipation
\label{sec.5.2}}

Various differents approached are known, aimed at getting dissipation ---and possibly
decoherence--- within quantum mechanics.  First of all, the simple introduction of
a ``chronon" (see, e.g., Refs.\cite{Caldirola1,CaldirolaM,Ruy1})
allows one to go on from differential to finite-difference equations, and in particular
to write down the quantum theoretical equations (Schroedinger's, Liouville-von
Neumann's, etc.) in three different ways: symmetrical, retarded, and andvanced.
The retarded ``Schroedinger" equation describes in a rather simple and natural way a dissipative
system, which exchanges (loses) energy with the environment. The corresponding non-unitary
time-evolution operator obeys a semigroup law and refers to irreversible
processes.
The retarded approach furnishes, moreover, an interesting way for solving the ``measurement
problem" in quantum mechanics, without any need for a wave-function collapse: see
Refs.\cite{Bonifacio1,Bonifacio2,Ghirardi,Ruy2,Ruy1}. The chronon theory can be
regarded as a peculiar ``coarse grained" description of the time evolution.

Let us stress that it has been shown that the
mentioned discrete appraoch can be replaced with a continuous one, at the price of
introducing a {\em non-hermitian} Hamiltonian: see, e.g., Ref.\cite{CasagrandeMontaldi}.

Further relevant work can be found, for instance, in papers like
\cite{Mignani1,Caldirola2,Janussis,JanussisM1,JanussisM2},
and refs. therein.

Let us add, at this point, that much work is still needed for the
description of time irreversibility at the microscopic level.
Indeed, various approaches have been proposed, in which new
parameters are introduced (regulation or dissipation) into the
microscopic dynamics (building a bridge, in a sense, between
microscopic structure and macroscopic characteristics). Besides
the Caldirola-Kanai\cite{Caldirola.1941.NC,Kanai.1948.PTP} Hamiltonian
\begin{equation}
  \hat{H}_{CK} (t) =
  -\,\displaystyle\frac{\hbar^{2}}{2m}\:
  \displaystyle\frac{\partial^{2}}{\partial x^{2}} e^{-\gamma t} +
  V (x) \; e^{\gamma t}
\label{eq.2.6}
\end{equation}

\noindent
(which has been used, e.g., in Ref.\cite{Angelopoulon}), other rather simple approaches, based of
course on the Schr\"{o}dinger equation
\begin{equation}
  i\hbar\, \displaystyle\frac{\partial}{\partial t}\; \Psi(x,t) =  \hat{H}\, \Psi(x,t) \, ,
\label{eq.2.1}
\end{equation}
and adopting a microscopic dissipation defined via a coefficient
of extinction $\gamma$, are for instance the following ones:

A) {\em Non-linear (non-hermitian) Hamiltonians}
\begin{equation}
  \hat{H}_{nl} =
  -\:\displaystyle\frac{\hbar^{2}}{2m}\:
  \displaystyle\frac{\partial^{2}}{\partial x^{2}} + V(x) + \hat{W} \; ,
\label{eq.2.2}
\end{equation}
 with ``potential'' operators $\hat{W}$ of the type:

\begin{enumerate}
\item
Kostin's operator (see Ref.\cite{Kostin.1972.JCP}):
\begin{equation}
  \hat{W}_{K} =
  -\:\displaystyle\frac{i\hbar}{2m}\:
  \biggl\{
    \displaystyle\frac{\ln{\Psi}}{\Psi^{*}} -
    \biggl\langle \displaystyle\frac{\ln{\Psi}}{\Psi^{*}} \biggr\rangle
  \biggr\} \; ;
\label{eq.2.3}
\end{equation}

\item
Albrecht's operator (see Ref.\cite{Albrecht.1975.PLB}):
\begin{equation}
  \hat{W}_{A} (x) = \langle p\, \rangle\, (x - \langle x\rangle) \; ,
\label{eq.2.4}
\end{equation}
where $\langle\,\rangle$ is the averaging produced over $| \Psi (x) |^{2}$;

\item
Ref.\cite{Hasse.1975.JMP}:
\begin{equation}
  \hat{W}_{H} (x) = \displaystyle\frac{1}{4}\: \Bigl[x - \langle x\rangle,\; p + \langle p\,\rangle\Bigr]_{+} \, ,
\label{eq.2.5}
\end{equation}
where $[A,B]_{+}$ is the anticommutator: $[A,B]_{+} = AB+BA$.

\end{enumerate}

B) {\em Linear (non-hermitian) Hamiltonians:}

\begin{enumerate}
\item
Ref.\cite{Gisin.1982.PA}:
\begin{equation}
  \hat{H}_{G} =
  (1 - i\gamma)\, \hat{H} + i\gamma \langle \hat{H} \rangle \; ;
\label{eq.2.7}
\end{equation}

\item
Ref.\cite{Exner.1983.JMP}:
\begin{equation}
  \hat{H}_{E} =
  \hat{H} + i\,\hat{W}(x)- i\, \langle \hat{W} (x)\, \rangle \; .
\label{eq.2.7}
\end{equation}
\end{enumerate}

One might recall also the important, so-called ``microscopic
models"\cite{Leggett}, even if they are not based on the
Schroedinger equation.

All such proposals are to be further investigated, and completed, since they
have not been apparently exploited enough, till now. Let us remark, just as
an example, that it would be desirable to take into deeper consideration other
related phenomena, like the ones associated with the ``Hartman effect"
(and ``generalized Hartman effect")
\cite{PhysRep2004,Olkhovsky_Recami.1992PR-1995JP-2004PR-2007IJMP,2bar1,Aharonov,2bar2,JMO},
in the case of tunneling with
dissipation: a topic faced in few papers, like \cite{RacitiSalesi,NimtzDiss}.

As a small contribution of ours, in the Appendix we present a scheme
of iterations (successive approximations) as a possible tool for
explicit calculations of wave-functions in the presence of
dissipation, by using as an example the simple Albreht's
potential. Our scheme may be useful, in any case, for the
investigation of possible violations of the Hartman effect, as
well as for analyzing a few irreversible phenomena. See the Appendix.

At last, let us incidentally recall that two generalized
Schroedinger equations, introduced by
Caldirola\cite{Caldirola2,Caldirola1976} in order to
describe two different dissipative processes (behavior of open
systems, and the radiation of a charged particle) have been shown
---see, e.g., Ref.\cite{Mignani83})--- to possess the same
algebraic structure of the Lie-admissible type\cite{Santilli}.

% *******************************************************************************************************************

% *******************************************************************************************************************
% \newpage
\section{Some conclusions
\label{sec.6}}

1. We have shown that the Time operator (\ref{eq.2.1.1}), hermitian even if non-selfadjoint, works for any quantum
collisions or motions, in the case
of a continuum energy spectrum, in non-relativistic quantum mechanics and in one-dimensional quantum electrodynamics.
The uniqueness of the (maximal) hermitian time operator (\ref{eq.2.1.1}) directly follows from the uniqueness of
the Fourier-transformations from the time to the energy representation. The time operator
(\ref{eq.2.1.1}) has been fruitfully used in the case, for instance, of tunnelling times (see
Refs.[18-21]), and of nuclear reactions and decays (see Refs.[7,8] and also
\cite{Olkhovsky.1992EL-1992NPA-1993NPA,Olkhovsky.2006.CEJP}). We have discussed the advantages of such an approach
with respect to POVM's, which is not applicable for three-dimensional particle collisions, within a wide
class of Hamiltonians.

The mathematical properties of the present Time operator have actually demonstrated
---without introducing any new physical postulates--- that {\em time} can be regarded as a quantum-mechanical observable,
at the same degree of other physical quantities (energy, momentum, spatial coordinates,...).

The commutation relations (eqs.(8),(22),(31)) here analyzed, and the uncertainty relations (\ref{eq.2.1.15}), result
to be analogous to those known for other pairs of canonically conjugate observables (as for coordinate
$\hat{x}$ and momentum $\hat{p}_{x}$, in the case of Eq.(\ref{eq.2.1.15})). Of course, our new relations do not
replace, but merely extend the meaning of the classic time and energy uncertainties, given e.g. in Ref.[49].

In subsection 2.6, we have studied the properties of Time, as an observable, for quantum-mechanical systems with
{\em discrete} energy spectra.

2. Let us recall that the Time operator (\ref{eq.2.1.1}), and relations (\ref{eq.2.1.2}),
(\ref{eq.2.1.3}), (\ref{eq.2.1.4}), (\ref{eq.2.3.4}), (\ref{eq.2.3.5}),
have been used for the temporal analyzis of nuclear reactions and decays
in Refs.[7,8]; as well as of new phenomena, about time delays-advances in nuclear physics,
in Refs.\cite{Olkhovsky.1992EL-1992NPA-1993NPA}, and about time
resonances or explosions of highly excited compound nuclei, in Refs.\cite{Olkhovsky.2006.CEJP}. \
Let us also recall that, besides the time operator, other quantities, to which (maximal) hermitian
operators correspond, can be analogously regarded as quantum-physical observables: For example, von Neumann
himself\cite{VonNeumann.1955,Recami.1976)} considered the case of the momentum operator $-i\partial / \partial x$ in a
semi-space with a rigid wall orthogonal to the $x$-axis at $x=0$, or of the radial momentum $-i \partial / \partial r$,
even if both act on packets defined only over the positive $x$ or $r$ axis, respectively.

Subsection 2.5 has been devoted to a new ``hamiltonian approach": namely, to the introduction of the analogue of the
``Hamiltonian" for the case of the Time operator.

3. In Section \ref{sec.3}, we have proposed a suitable generalization for the Time operator (or, rather, for a
Space-Time operator) in relativistic quantum mechanics. For instance, for the
Klein-Gordon case, we have shown that the hermitian part of the three-position operator $\hat{x}$ is nothing but
the Newton-Wigner operator, and corresponds to a point-like position; while its anti-hermitian part can be regarded
as yielding the sizes of an extended-type (ellipsoidal) localization. When dealing with a 4-position operator, one
finds that the Time operator is selfadjoint for unbounded energy spectra, while it is a (maximal) hermitian operator
when the kinetic energy, and the total energy, are bounded from below, as for a free particle.  We have extensively made
recourse, in the latter case, to {\it bilinear} forms, which dispense with the necessity of eliminating the lower point
---corresponding to zero velocity--- of the spectra. It would be interesting to proceed to further generalizations
of the 3- and 4-position operator for other relativistic cases, and analyze the localization problems associated with
Dirac particles, or in 2D and 3D quantum electrodynamics, etc.  Work is in progress on time analyses in 2D
quantum electrodynamic, for application, e.g., to frustrated (almost total) internal reflections. Further work has
still to be done also about the joint consideration of particles and antiparticles.

4. Section 4 has been devoted to the association of unstable states (decaying "resonances") with the
eigenvectors of quasi-hermitian[40,41,47] Hamiltonians.
% One more direction of application of
%non-hermitian quantum dynamics had been earlier proposed and
%afterwards partially developed for the analysis of the
%decaying states evolution with complex energies, and also for calculating
%those complex energies on the basis of suitable couplings between the
%discrete states of unperturbed subsystem and energy-conserving
%continuum [47].

5. Non-hermitian Hamiltonians, and non-unitary time-evolution
operators, can play an important role also in microscopic quantum dissipation[59--71]: namely, in
getting decoherence through interaction with the enviroment[61,62]. This topic is touched
in Section 5; together with questions related with collisions in absorbing
media. In particular, in Sec.5 we mention also the case of the optical model in
nuclear physics; without forgettig that non-hermitian operators show up even in the case
of tunnelling ---e.g., in fission phenomena--- with quantum dissipation, and of quantum friction.
As to the former topic of microscopic quantum dissipation, among the many approaches to quantum
irreversibility we have discussed in Sec.5.2 a possible solution of the quantum
measurement problem (via interaction with the environment) by the introduction
of finite-difference equations (e.g., in terms of a ``chronon").

6. Let us eventually observe that the ``dual equations" \ref{eq.2.5.6}) and (\ref{eq.2.5.7})
seem to be promising also for the study the initial stage of our cosmos,
when tunnellings can take place through the barriers which appear in quantum gravity in the limiting case of
quasi-Schr\"{o}edinger equations.
% *******************************************************************************************************************

% *******************************************************************************************************************
\section{Acknowledgements
\label{sec.7}}

Part of this paper is based on work performed by one of us in
collaboration with P.Smrz, and with A.Agodi, M.Baldo and
A.Pennisi di Floristella. Thanks are moreover due for stimulating
discussions to Y.Aharonov, A.S.Holevo, V.L.Lyuboshitz, R.Mignani, V.Petrillo,
G.Salesi, B.N.Zakhariev, and M.Zamboni-Rached.
% *******************************************************************************************************************

\

\

\

\

\

\

% \newpage
\section{APPENDIX
\label{sec.8}}
\centerline {\bf TIME-DEPENDENT SCHR\"ODINGER EQUATION WITH DISSIPATIVE TERMS}

\

\

\subsection{Introduction
\label{sec.8.1}}

Let's consider the time-dependent Schr\"{o}dinger equation:
\begin{equation}
  i \displaystyle\frac{\partial}{\partial t} \Psi(x,t) =
  \biggl(-\displaystyle\frac{\partial^{2}}{\partial x^{2}} + V(x,t) \biggr) \:
  \Psi(x,t),
\label{eq.2.1}
\end{equation}
where we put $\hbar = 1$.
Let us rewrite the time-dependent wave function (WF), $\Psi(x,t)$ (which can be considered as
a wave-packet (WP)), in the form of a Fourier integral:
\begin{equation}
  \Psi(x,t) =
  \displaystyle\int\limits_{0}^{E_{0}}
  g(E) \, e^{-iEt} \, \varphi(E,x) \: dE \, ,
\label{eq.2.2}
\end{equation}
where $\varphi(E,x)$ is the WF component independent of time,
and $g(E)$ is a weight factor.
One can choose the function $g(E)$ to be, e.g., a \emph{Gaussian}:
\begin{equation}
  g(E) = A \: e^{-a^2 (k-\bar{k})^2} \, .
\label{eq.2.3}
\end{equation}
Here, $A$ and $a$ are constants, and $\bar{k}$ is the selected value for the impulse,
constituting the center of the WP.
We substitute the Fourier-expansion (\ref{eq.2.2}) of WF into eq.(\ref{eq.2.1}). Thus,
the l.h.s. of this equation trasforms into

\begin{equation}
  i \displaystyle {\frac{\partial}{\partial t} \Psi(x,t) =
  \int\limits_{0}^{E_{0}}
    g(E) \, e^{-iEt} \, \varphi(E,x) \: E dE} \, .
\label{eq.2.5}
\end{equation}
Afterwards, the r.h.s. of eq.(\ref{eq.2.1}) gets transformed into
\begin{equation}
\begin{array}{ccl}
  \biggl(-\displaystyle\frac{\partial^{2}}{\partial x^{2}} + V(x,t) \biggr)
  \Psi(x,t) =
  -\displaystyle\int\limits_{0}^{E_{0}}
    g(E) e^{-iEt}
    \displaystyle\frac{\partial^{2} \varphi(E,x) }{\partial x^{2}} \: dE +
  \displaystyle\int\limits_{0}^{E_{0}}
    g(E) V(x, \bar{E}, t) e^{-iEt} \varphi(E,x) \: dE.
\end{array}
\label{eq.2.6}
\end{equation}
Therefore, the whole equation (\ref{eq.2.1}) has been transformed into
\begin{equation}
  \displaystyle\int\limits_{0}^{E_{0}}
    g(E) \, e^{-iEt} \, \varphi(E,x) \: E dE =
  -\displaystyle\int\limits_{0}^{E_{0}}
    g(E) \, e^{-iEt}
    \displaystyle\frac{\partial^{2} \varphi(E,x) }{\partial x^{2}} \: dE +
  \displaystyle\int\limits_{0}^{E_{0}}
    g(E) \, V(x, \bar{E}, t) \, e^{-iEt} \, \varphi(E,x) \: dE.
\label{eq.2.7}
\end{equation}

Let us now apply the inverse Fourier-transformation to this equation. Its left part
becomes
\begin{equation}
\begin{array}{ccl}
  \displaystyle\frac{1}{2\pi}
  \displaystyle\int dt \: e^{iE't}
  \displaystyle\int\limits_{0}^{E_{0}}
    g(E) \, e^{-iEt} \, \varphi(E,x) \: E dE =

  \displaystyle\frac{1}{2\pi}
  \displaystyle\int\limits_{0}^{E_{0}}
    dE \: E \, g(E) \, \varphi(E,x)
  \displaystyle\int e^{i(E'-E)t} dt = \\

  = \displaystyle\int\limits_{0}^{E_{0}}
    g(E) \, \varphi(E,x) \delta(E'-E) \: E dE =
  g(E') \, E' \, \varphi(E',x) \, ;
\end{array}
\label{eq.2.8}
\end{equation}
while its right part becomes
\begin{equation}
\begin{array}{ccl}
  -\displaystyle\frac{1}{2\pi}
  \displaystyle\int dt \: e^{iE't}
  \displaystyle\int\limits_{0}^{E_{0}}
    g(E) \, e^{-iEt}
    \displaystyle\frac{\partial^{2} \varphi(E,x) }{\partial x^{2}} \: dE +
  \displaystyle\frac{1}{2\pi}
  \displaystyle\int dt \: e^{iE't}
  \displaystyle\int\limits_{0}^{E_{0}}
    g(E) \, V(x, \bar{E}, t) \, e^{-iEt} \, \varphi(E,x) \: dE = \\

  = -\displaystyle\frac{1}{2\pi}
  \displaystyle\int\limits_{0}^{E_{0}}
    dE \, g(E) \displaystyle\frac{\partial^{2} \varphi(E,x) }{\partial x^{2}}
  \displaystyle\int e^{i(E'-E)t} \: dt +

  \displaystyle\frac{1}{2\pi}
  \displaystyle\int\limits_{0}^{E_{0}}
    dE \, g(E) \, \varphi(E,x)
  \displaystyle\int V(x,\bar{E},t) \, e^{i(E'-E)-t} \: dt = \\

  = - g(E') \displaystyle\frac{\partial^{2} \varphi(E',x) }{\partial x^{2}} +
  \displaystyle\frac{1}{2\pi}
  \displaystyle\int\limits_{0}^{E_{0}}
    dE \, g(E) \, \varphi(E,x)
  \displaystyle\int V(x,\bar{E},t) \, e^{i(E'-E)t} \: dt.
\end{array}
\label{eq.2.9}
\end{equation}
As a result, we obtain eq.(\ref{eq.2.7}) in the form
\begin{equation}
\begin{array}{ccl}
  g(E') \, E' \, \varphi(E',x) =
  - g(E') \displaystyle\frac{\partial^{2} \varphi(E',x) }{\partial x^{2}} +
  \displaystyle\frac{1}{2\pi}
  \displaystyle\int\limits_{0}^{E_{0}}
    dE \, g(E) \, \varphi(E,x)
  \displaystyle\int V(x,\bar{E},t) \, e^{i(E'-E)t} \: dt.
\end{array}
\label{eq.2.10}
\end{equation}
% ***************************************************************************

% ***************************************************************************
\subsection{The case of the simple Albreht's potential
\label{sec.8.2}}

Just as an example of a possible potential $V(x,t)$, let us choose
\begin{equation}
  V(x,t) = V_{0}(x) + \gamma \; W_{A}(x) \, .
\label{eq.3.1}
\end{equation}
where $W_{A}(x)$ is the simple Albreht's dissipation term.
Here, $\gamma$ is a constant, $V_{0}(x)$ is the usual stationary component of $V(x)$, and
the dissipative component of $V(x)$ has the form
\begin{equation}
  W_{A}(x) = <p> (x - <x>),
\label{eq.3.2}
\end{equation}
where the averages are fulfilled by integrating over $x$ by means of the functions
$\Psi^{*}(x,t)$ and $\Psi(x,t)$. For the right part of eq.(\ref{eq.3.2}) one gets
\begin{equation}
\begin{array}{ccl}
  <p> & = &
  -i \displaystyle\int dx
    \displaystyle\int\limits_{0}^{E_{0}} dE_{1}
    \displaystyle\int\limits_{0}^{E_{0}} dE_{2} \:
      g(E_{1}) \, g(E_{2}) \:
      e^{i(E_{1}-E_{2})t} \:
      \varphi^{*}(E_{1},x) \,
      \displaystyle\frac{\partial \varphi(E_{2},x)}{\partial x}, \\

  <x> & = &
  \displaystyle\int dx
    \displaystyle\int\limits_{0}^{E_{0}} dE_{3}
    \displaystyle\int\limits_{0}^{E_{0}} dE_{4} \:
      g(E_{3}) \, g(E_{4}) \:
      e^{i(E_{3}-E_{4})t} \:
      x \:
      \varphi^{*}(E_{3},x) \,
      \varphi(E_{4},x) \, ;
\end{array}
\label{eq.3.3}
\end{equation}
so that the total potential $V(x,t)$ becomes
\begin{equation}
\begin{array}{ccl}
  V(x,t) & = &
  V_{0}(x) -
    i\gamma \:
    \displaystyle\int dx_{1}
    \displaystyle\int dx_{2}
    \displaystyle\int\limits_{0}^{E_{0}} dE_{1}
    \displaystyle\int\limits_{0}^{E_{0}} dE_{2}
    \displaystyle\int\limits_{0}^{E_{0}} dE_{3}
    \displaystyle\int\limits_{0}^{E_{0}} dE_{4} \;

      g(E_{1}) \, g(E_{2}) \, g(E_{3}) \, g(E_{4}) \\

  & \times &
      e^{i(E_{1}-E_{2}+E_{3}-E_{4})t} \:
      \Bigl( x - x_{2} \Bigr) \:
      \varphi^{*}(E_{1},x_{1}) \,
      \displaystyle\frac{\partial \varphi(E_{2},x_{1})}{\partial x_{1}} \,
      \varphi^{*}(E_{3},x_{2}) \,
      \varphi(E_{4},x_{2}).
\end{array}
\label{eq.3.5}
\end{equation}

Taking into account this, we find the second term, in the r.h.s. of eq.(\ref{eq.2.10}),
to be:
\begin{equation}
\begin{array}{l}
  \displaystyle\frac{1}{2\pi}
  \displaystyle\int\limits_{0}^{E_{0}}
    dE \, g(E) \, \varphi(E,x)
  \displaystyle\int V(x,\bar{E},t) \, e^{i(E'-E)t} \: dt = \\

  = \;
  \displaystyle\int\limits_{0}^{E_{0}}
    dE \, g(E) \, \varphi(E,x) \; V_{0}(x) \: \delta (E'-E) \: - \\

  - \;
  i \gamma \displaystyle\int\limits_{0}^{E_{0}}
    dE \, g(E) \, \varphi(E,x) \;
    \displaystyle\int dx_{1}
    \displaystyle\int dx_{2}
    \displaystyle\int\limits_{0}^{E_{0}} dE_{1}
    \displaystyle\int\limits_{0}^{E_{0}} dE_{2}
    \displaystyle\int\limits_{0}^{E_{0}} dE_{3}
    \displaystyle\int\limits_{0}^{E_{0}} dE_{4} \:
      g(E_{1}) \, g(E_{2}) \, g(E_{3}) \, g(E_{4}) \\

  \times \:
      \Bigl( x - x_{2} \Bigr) \:
      \varphi^{*}(E_{1},x_{1}) \,
      \displaystyle\frac{\partial \varphi(E_{2},x_{1})}{\partial x_{1}} \,
      \varphi^{*}(E_{3},x_{2}) \,
      \varphi(E_{4},x_{2}) \;
      \delta(E'-E+E_{1}-E_{2}+E_{3}-E_{4}) = \\

      \

  = \; g(E') \, \varphi(E',x) \, V_{0}(x) \: - \\

  - \; i \gamma \;
    \displaystyle\int dx_{1}
    \displaystyle\int dx_{2}
    \displaystyle\int\limits_{0}^{E_{0}} dE_{1}
    \displaystyle\int\limits_{0}^{E_{0}} dE_{2}
    \displaystyle\int\limits_{0}^{E_{0}} dE_{3}
    \displaystyle\int\limits_{0}^{E_{0}} dE_{4} \:
      g(E_{1}) \, g(E_{2}) \, g(E_{3}) \, g(E_{4}) \, g(E'') \\

  \times \:
      \Bigl( x - x_{2} \Bigr) \:
      \varphi^{*}(E_{1},x_{1}) \,
      \displaystyle\frac{\partial \varphi(E_{2},x_{1})}{\partial x_{1}} \,
      \varphi^{*}(E_{3},x_{2}) \,
      \varphi(E_{4},x_{2}) \,
      \varphi(E'',x),
\end{array}
\label{eq.3.7}
\end{equation}
where
\begin{equation}
  E'' = E' + E_{1} - E_{2} + E_{3} - E_{4}.
\label{eq.3.8}
\end{equation}
As a consequence, the whole eq.~(\ref{eq.2.10}) gets transformed into
\[
\begin{array}{ccl}
  g(E') \, E' \, \varphi(E',x) =
  - g(E') \displaystyle\frac{\partial^{2} \varphi(E',x) }{\partial x^{2}} +

  g(E') \, \varphi(E',x) \, V_{0}(x) \\

  - \; i \gamma \
    \displaystyle\int dx_{1}
    \displaystyle\int dx_{2}
    \displaystyle\int\limits_{0}^{E_{0}} dE_{1}
    \displaystyle\int\limits_{0}^{E_{0}} dE_{2}
    \displaystyle\int\limits_{0}^{E_{0}} dE_{3}
    \displaystyle\int\limits_{0}^{E_{0}} dE_{4} \:
      g(E_{1}) \, g(E_{2}) \, g(E_{3}) \, g(E_{4}) \, g(E'') \\

  \times \:
      \Bigl( x - x_{2} \Bigr) \:
      \varphi^{*}(E_{1},x_{1}) \,
      \displaystyle\frac{\partial \varphi(E_{2},x_{1})}{\partial x_{1}} \,
      \varphi^{*}(E_{3},x_{2}) \,
      \varphi(E_{4},x_{2}) \,
      \varphi(E'',x)
\end{array}
\]
or (with the change of variables $E' \to E$)
\begin{equation}
\begin{array}{ccl}
  \biggl(
    - \displaystyle\frac{\partial^{2}}{\partial x^{2}} +
    V_{0}(x) - E
  \biggr) \, \varphi(E,x) = \\

  = i \gamma \:
    \displaystyle\int dx_{1}
    \displaystyle\int dx_{2}
    \displaystyle\int\limits_{0}^{E_{0}} dE_{1}
    \displaystyle\int\limits_{0}^{E_{0}} dE_{2}
    \displaystyle\int\limits_{0}^{E_{0}} dE_{3}
    \displaystyle\int\limits_{0}^{E_{0}} dE_{4} \:
      \displaystyle\frac
        {g(E_{1}) \, g(E_{2}) \, g(E_{3}) \, g(E_{4}) \, g(E'')}
        {g(E)} \\

  \times \:
      \Bigl( x - x_{2} \Bigr) \:
      \varphi^{*}(E_{1},x_{1}) \,
      \displaystyle\frac{\partial \varphi(E_{2},x_{1})}{\partial x_{1}} \,
      \varphi^{*}(E_{3},x_{2}) \,
      \varphi(E_{4},x_{2}) \:
      \varphi(E'',x) \, .
\end{array}
\label{eq.3.9}
\end{equation}
We have thus obtained for this case the time-independent Schr\"oedinger equation, by taking
however into account dissipation via the parameter $\gamma$. \ Of course, when $\gamma$ tends
to zero, one goes back to the stationary Schr\"{o}dinger equation.
% ***************************************************************************

% ***************************************************************************
\subsection{Method of the successive approximations
\label{sec.8.3}}

Assuming the coefficient $\gamma$ to be small, one can find the unknown
function $\varphi (x)$ in the simplified form
\begin{equation}
  \varphi (x) = \varphi_{0} (x) + \gamma \; \varphi_{1} (x),
\label{eq.4.1}
\end{equation}
where as function $\varphi_{0}(x)$ it has been used the standard WF of the time-independent
Schr\"{o}dinger equation with potential $V_{0}(x)$ and energy $E_{0}$:
\begin{equation}
  \biggl( -\displaystyle\frac{\partial^{2}}{\partial x^{2}} + V_{0}(x) \biggr) \varphi_{0} (x) =
  E_{0}\, \varphi_{0} (x).
\label{eq.4.2}
\end{equation}
Substituting solution (\ref{eq.4.1}) into eq.(\ref{eq.3.9}), we obtain a new equation containing
all the powers $n$ of $\gamma$, namely, the $\gamma^{n}$.
Let us confine ourselves, however, to write down this equation with accuracy up to
$\gamma^{1}$ only:
\begin{equation}
\begin{array}{ccl}
  \biggl(
    - \displaystyle\frac{\partial^{2}}{\partial x^{2}} +
    V_{0}(x) - E
  \biggr) \,
  \Bigl( \varphi_{0} (E,x) + \gamma\varphi_{1} (E,x) \Bigr) = \\

  = i \gamma \;    \displaystyle\int dx_{1}
    \displaystyle\int dx_{2}
    \displaystyle\int\limits_{0}^{E_{0}} dE_{1}
    \displaystyle\int\limits_{0}^{E_{0}} dE_{2}
    \displaystyle\int\limits_{0}^{E_{0}} dE_{3}
    \displaystyle\int\limits_{0}^{E_{0}} dE_{4} \:
      \displaystyle\frac
        {g(E_{1}) \, g(E_{2}) \, g(E_{3}) \, g(E_{4}) \, g(E'')}
        {g(E)} \\

  \times \:
      \Bigl( x - x_{2} \Bigr) \:
      \varphi_{0}^{*}(E_{1},x_{1}) \,
      \displaystyle\frac{\partial \varphi_{0}(E_{2},x_{1})}{\partial x_{1}} \,
      \varphi_{0}^{*}(E_{3},x_{2}) \,
      \varphi_{0}(E_{4},x_{2}) \:
      \varphi_{0}(E'',x) \, ,
\end{array}
\label{eq.4.3}
\end{equation}
where the unknown $\varphi_{1}(x)$ does not appear any longer, of course,into the r.h.s. of
this equation.

Taking
\begin{equation}
  E = E_{0} \, ,
\label{eq.4.4}
\end{equation}
we can rewrite in eq.(\ref{eq.4.3}), separately, the various terms with different powers
of $\gamma$. When limiting ourselves to $n=0,1$, we obtain
\begin{equation}
\begin{array}{cc}
  \gamma^{0}: &
  \biggl(
    - \displaystyle\frac{\partial^{2}}{\partial x^{2}} + V_{0}(x) - E_{0}
  \biggr) \,
  \varphi_{0} (E_{0},x) = 0, \\

  \gamma^{1}: &
  \biggl(
    - \displaystyle\frac{\partial^{2}}{\partial x^{2}} + V_{0}(x) - E_{0}
  \biggr) \,
  \varphi_{1} (E_{0},x) = \\

  & = i \gamma \;
    \displaystyle\int dx_{1}
    \displaystyle\int dx_{2}
    \displaystyle\int\limits_{0}^{E_{0}} dE_{1}
    \displaystyle\int\limits_{0}^{E_{0}} dE_{2}
    \displaystyle\int\limits_{0}^{E_{0}} dE_{3}
    \displaystyle\int\limits_{0}^{E_{0}} dE_{4} \:
      \displaystyle\frac
        {g(E_{1}) \, g(E_{2}) \, g(E_{3}) \, g(E_{4}) \, g(E'')}
        {g(E_{0})} \\

  & \times \:
      \Bigl( x - x_{2} \Bigr) \:
      \varphi_{0}^{*}(E_{1},x_{1}) \,
      \displaystyle\frac{\partial \varphi_{0}(E_{2},x_{1})}{\partial x_{1}} \,
      \varphi_{0}^{*}(E_{3},x_{2}) \,
      \varphi_{0}(E_{4},x_{2}) \:
      \varphi_{0}(E'',x),
\end{array}
\label{eq.4.5}
\end{equation}
where
\begin{equation}
  E'' = E_{0} + E_{1} - E_{2} + E_{3} - E_{4}.
\label{eq.4.6}
\end{equation}
The first equation holds when dissipation is absent. The second equation determines
the unknown function $\varphi_{1}$ in terms of the given $\varphi_{0}$: It results
to be an ordinary differential equation of the second order, that can be solved by
the ordinary numerical methods; but we deem convenient, here, to skip any
numerical evaluations.

Let us just observe that, of course, one can go on iteratively to higher values of $n$.
% ***************************************************************************

% ***************************************************************************
\newpage
% \section*{References} % needed only for style of Rep.Prog.Phys.
% \bibliography{OpNSA_v3}
%       style article
%

% \bibitem{Weisskopf_Wigner.1930} % 45 (37)
%   V.~F.~Weisskopf and E.~P.~Wigner,
%   Z.~Phys. \textbf{63}, 54 (1930).

% \bibitem{Feshbach.1958.AP} % 46 (38)
%   H.~Feshbach,
%   Ann.~Phys. \textbf{5}, 357 (1958).

% \bibitem{Backer.1983PRL-1984PRA} % 47 (39)
%   H.~C.~Backer,
%   Phys.~Rev.~Lett. \textbf{50}, 1579 (1983);
%   Phys.~Rev. \textbf{A30}, 773 (1984).

% \bibitem{Baker.1990.PRA} % 48 (40)
%   H.~C.~Baker and R.~L.~Singleton,
%   Phys.~Rev. \textbf{A42}, 10 (1990).
%-----------------------------------------------------------------------------------------------------------------------

\end{document}